\documentclass[conference,10pt]{IEEEtran}
\makeatletter
\def\ps@headings{%
\def\@oddhead{\mbox{}\scriptsize\rightmark \hfil \thepage}%
\def\@evenhead{\scriptsize\thepage \hfil \leftmark\mbox{}}%
\def\@oddfoot{}%
\def\@evenfoot{}}
\makeatother
\IEEEoverridecommandlockouts

\usepackage{amssymb}    % symboles ams
\usepackage{amsmath}    % symboles ams maths
\usepackage{amsfonts}
\usepackage{bbm}
\usepackage{here}
\usepackage{mathrsfs}
\usepackage{graphicx}
\usepackage{color}
\usepackage{psfrag}
\usepackage{epsfig}
\usepackage{comment}
\usepackage{multirow}
\usepackage{mathtools}
\usepackage{url}

\newcounter{lawc}
\newtheorem{result}{Result}[section]

\newtheorem{proposition}{Proposition}[section]
\newtheorem{define}{Definition}[section]

\newtheorem{example}{Example}[section]

\newtheorem{law}[lawc]{Law}
 {\everymath{\displaystyle\everymath{}}\equation}%
 {\endequation}

\DeclareMathOperator*{\col}{col}
\DeclareMathOperator{\Sq}{Sq}

\DeclareMathOperator{\eps}{\varepsilon}
\DeclareMathOperator{\RTT}{RTT}

\newcommand{\menlem}{\hfill \ensuremath{\blacktriangle}}
\newcommand{\mendlaw}{\hfill \ensuremath{\therefore}}
\newcommand{\mendprop}{\hfill \ensuremath{\triangledown}}

{\ensuremath{\left[\begin{array}{ccccccccccccccccccccccccccccccccccccccccccccccccccccccccccccccccccccc}}}{\ensuremath{\end{array}\right]}}

\renewcommand{\paragraph}[1]{\textbf{#1}}

\title{A conservation-law-based modular fluid-flow model for network congestion modeling}
\author{Corentin~Briat, Emre A.~Yavuz and Gunnar~Karlsson\thanks{C. Briat, E. A. Yavuz, and G. Karlsson are with ACCESS Linnaeus Centre at the Royal Institute of Technology (KTH), Stockholm, SWEDEN. Email: \{cbriat, emreya, gk\}@kth.se}\thanks{This work has been supported by the ACCESS project, KTH, Stockholm, Sweden. http://www.access.kth.se/}}

\begin{document}
\maketitle

\thispagestyle{empty}
\pagestyle{empty}

\begin{abstract}
A modular fluid-flow model for network congestion analysis and control is proposed. The model is derived from an information conservation law stating that the information is either in transit, lost or received. Mathematical models of network elements such as queues, users, and transmission channels, and network description variables, including sending/acknowledgement rates and delays, are inferred from this law and obtained by applying this principle locally. The modularity of the devised model makes it sufficiently generic to describe any network topology, and appealing for building simulators. Previous models in the literature are often not capable of capturing the transient behavior of the network precisely, making the resulting analysis inaccurate in practice. Those models can be recovered from exact reduction or approximation of this new model. An important aspect of this particular modeling approach is the introduction of new tight building blocks that implement mechanisms ignored by the existing ones, notably at the queue and user levels. Comparisons with packet-level simulations corroborate the proposed model.
\end{abstract}


\section{Introduction}\label{sec:introduction}
%\input{introduction}

%\section{Motivations}\label{sec:motivations}
%%\input{motivations}
%
%\section{Related works and contributions}\label{sec:related_work}
Network modeling is challenging due to the very heterogeneous nature of communication networks, mixing physics, electronics and computer science. This heterogeneity coupled with intrinsic properties of physical and mathematical laws prevent the development of an efficient bottom-up approach. This is why finding macroscopic laws capturing critical phenomena is of interest. These laws should provide an abstraction of the microscopic level by identifying and relating the fundamental macroscopic network parameters.

We derive a modular network model using three fundamental laws. The first law is a packet conservation law that facilitates the derivation of models as building blocks and simplifies their mathematical expression. The second law defines a model for queues and, finally, the last one concerns the existence of a user model. We will show that each law has implications for the network modeling problem and, more importantly, will allow to solve yet unresolved problems, especially at user level. One important property of the developed model is its modularity. Indeed, the modeling technique allows to develop each element independently of the others, leading then to \emph{building blocks} which may be interconnected as desired to build any network topology. The model shares exactly the same structure with a real network which is built by interconnection of several elements. This property is also very appealing for simulation purposes where a topology can be easily simulated by connecting the building-blocks.

The proposed model is developed in several steps. The first one is the modeling of lossless transmission channels directly from the first law. The second step is the derivation, again from the first law, of the so-called ACK-clocking model \cite{Jacobsson:08b}, which ties flow, flight-size and round-trip time (RTT) together. This result is of great importance in network modeling.

Based on the first and second laws, a causal RTT expression is developed in the third step. This causal RTT expression however requires an extension of the buffer model. Indeed, there are two main limitations to the buffer model usually considered in the literature (and as stated in the second law). First, the model does not explicitly define the queue as a FIFO queue (i.e. order preserving) in which the packets maintain their relative positions. An internal buffer description should capture this, at a flow level. To this aim, the flows should therefore be considered as very viscous repelling liquids which do not mix. The second limitation concerns the solving of the output flow separation problem, primordial for the description of buffer interconnections and, as we shall see later, for the derivation of an exact expression for the acknowledgment flows.

Finally, the last step is devoted to the derivation of a complete user model, based on the first and third laws. This part constitutes one important contributions of the paper. Indeed, the conversion of congestion window size into flow has been a major obstacle preventing the improvement of network models. The \emph{static-link model} \cite{Wang:05} assimilates the flows to be equal to the derivative of congestion window sizes. It has good modeling properties in the absence of cross-traffic and when propagation delays are homogeneous. It is however rather inaccurate in more realistic scenarios. This validity domain is theoretically proved in this paper by showing that the proposed model reduces to the static-link one when some conditions are met. The \emph{integrator link model} \cite{Vinnicombe:00b,Misra:00,Hollot:01,Low:02,Paganini:03b} improves the description by correcting the irrelevant behavior of the model when affected by cross-traffic. Yet, some characteristics of the buffer response were not well captured: the response speed and the high slope when the congestion window size increases. The \emph{joint-link model} \cite{Jacobsson:08, Moller:08} consisting of merging the static-link and integrator-link models has been introduced. This approach improves the network model by capturing some characteristics unmodeled by the previous descriptions. It has been shown that these flow models can, in fact, be considered as approximations of the ACK-clocking model \cite{Jacobsson:08, Moller:08,Tang:10} from which higher order approximations can also be defined. More recently, the ACK-clocking model has been exactly\footnote{Although the definitions for the ACK-clocking model slightly differ.} considered in \cite{Jacobsson:08,Tang:10} and has lead to important improvements in terms of precision. In this paper, we also do not make any approximations and use the ACK-clocking model in a new fashion, leading to a new user model explicitly using the received flow of acknowledgements. This new model exactly captures both the ACK-clocking and the decreasing of the congestion window size, the latter being not captured by the existing models.

The modeling technique is applied to a single-buffer/multiple-users topology for which it is possible to show that, under some certain conditions, the static-link model can be naturally recovered. This shows that the static-link model is more general than it was previously known \cite{Wang:05,Tang:10}. Unlike other versions of the static-link model, the obtained one involves a state-dependent time-delay \cite{Hartung:06} representing the queuing delay \cite{Briat:10}.

Finally, simulation results are provided for some simple topologies to compare the proposed model with previous ones. It is shown that the model reproduces the results of \cite{Jacobsson:08,Tang:10}, as expected since those models are based on the ACK-clocking model. By transitivity, the proposed model also matches NS-2 simulations and the experiments reported in \cite{Jacobsson:08,Tang:10}. Simulations also illustrate that the proposed model is also able to capture the behavior of the queues and users when the congestion window size decreases, as opposed to the existing ones. Another important property of the model is its modularity which makes it easy to manipulate when building network topologies and setting up simulations, where blocks just have to be connected to others.

\begin{figure}[H]
\begin{minipage}[b]{0.49\linewidth}
\centering
        \psfrag{u}[c][c]     {\footnotesize {User}}
        \psfrag{n}[c][c]     {\footnotesize {Network}}
        \psfrag{p}[c][c]     {\scriptsize{\shortstack{Sending\\flow}}}
        \psfrag{a}[c][c]     {\scriptsize{\shortstack{Congestion\\price}}}
  \includegraphics[width=0.7\textwidth]{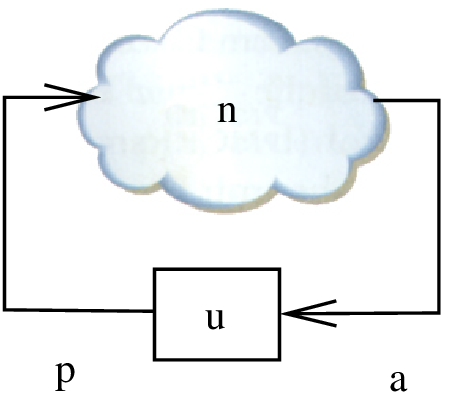}
  \caption{Usual abstract representation of congestion control}\label{fig:abscong1}
\end{minipage}
\hfill
\begin{minipage}[b]{0.49\linewidth}
  \centering
        \psfrag{u}[c][c]     {\footnotesize {User}}
        \psfrag{n}[c][c]     {\footnotesize {Network}}
        \psfrag{p}[c][c]     {\scriptsize{\shortstack{Sending\\flow}}}
%        \psfrag{a}[c][c]     {\scriptsize{\shortstack{Congestion\\measure\ }\shortstack{$+$\\\ }\shortstack{\ ACK\\flow}}}
        \psfrag{a}[c][c]     {\scriptsize{\shortstack{Congestion price\\$+$\\ACK flow}}}
  \includegraphics[width=0.7\textwidth]{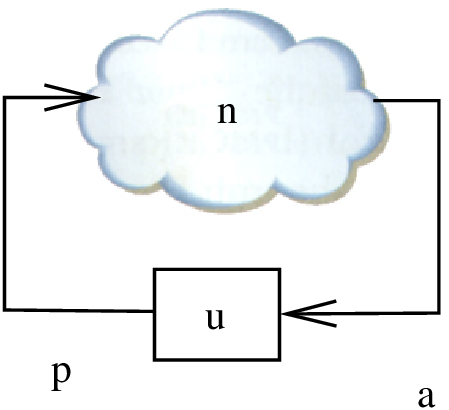}
  \caption{Considered abstract representation of congestion control}\label{fig:abscong2}
\end{minipage}
\end{figure} 

\section{Definitions and laws}\label{sec:preliminaries}
\subsection{Networks and Graphs}\label{sec:netgraph}

%Before entering in the modeling problem, it is convenient to define here some notations and introduce unusual graphs for network graphical representation. Let us recall first that, usually, a
%network can be viewed as a graph $\mathcal{G}:=(\mathcal{N}, \mathcal{L})$. The nodes set $\mathcal{N}$ can be divided in two families $\{u_i\}_i$ and $\{b_i\}_i$, the users and buffers respectively.

%A network graph is, in general, undirected since if an element $x$ can transmit data to the element $y$ then the inverse is also possible. We will consider however, in the current paper, that once connections and connecting circuits are established, the graph becomes directed. That is, it will be assumed that the transmission circuits do not change over time.

It is convenient to introduce here the particular network graph representation considered in the paper. It is different from the regular ones since it places all network elements on graph edges, leaving nodes with the role of connecting points, as in electrical circuits. Four types of nodes are distinguished: the input nodes $u_i^-$, $b_j^-$ and output nodes $u_i^+$, $b_j^+$ for user $u_i$ and buffer $b_j$, respectively. The superscripts have to be understood as a temporal order of reaction or causality: the data come at (-) and leave at (+). We denote any edge $E$ of the graph by $\langle x,y\rangle$ where $x$ and $y$ are the input and output nodes respectively. Moreover, given any edge $E$, the input and output nodes are given by $\beta(E)$ and $\eps(E)$ respectively. %We say that a point $x\in E$, \emph{the location $x$ lies on the edge $E$}, if there exists $\theta\in[0,1]$ such $x=(\eps(E)-\beta(E))\theta+\beta(E)$. Hence, we see any edge $E$ as a straight line between $\beta(E)$ and $\eps(E)$.

\begin{figure}
  \centering
        \psfrag{be}[c][c]     {{{$\langle b^-,b^+\rangle$}}}
        \psfrag{bi-}[c][c]     {{{$b^-$}}}
        \psfrag{bi+}[c][c]     {{{$b^+$}}}
        \psfrag{ui-}[c][c]     {{{$u^-$}}}
        \psfrag{ui+}[c][c]     {{{$u^+$}}}
        \psfrag{ue}[c][c]     {{{$\langle u^-,u^+\rangle$}}}
        \psfrag{l1e}[c][c]     {{{$\langle u^+,b^-\rangle$}}}
        \psfrag{l2e}[c][c]     {{{$\langle b^+,u^-\rangle$}}}
        \psfrag{b}[c][c]     {{{$b$}}}
        \psfrag{l1}[c][c]     {{{$\ell_1$}}}
        \psfrag{l2}[c][c]     {{{$\ell_2$}}}
        \psfrag{u}[c][c]     {{{$u$}}}
  \includegraphics[width=0.25\textwidth]{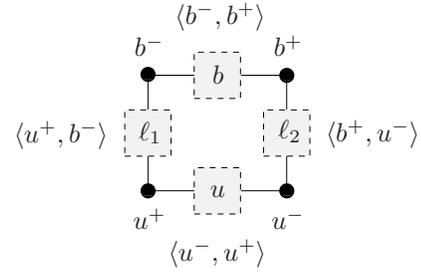}
  \caption{Example of graph with 4 edges: one user edge $u=\langle u^-,u^+\rangle$, one buffer edge $b=\langle b^-,b^+\rangle$ and two transmission edges $\ell_1=\langle u^+,b^-\rangle$, $\ell_2=\langle b^+,u^-\rangle$}\label{fig:graph}
\end{figure}

According to these definitions, a queue edge is always denoted by $\langle b_i^-,b_i^+\rangle$, a user edge by $\langle u_i^-,u_i^+\rangle$ and a transmission edge by $\langle b_i^+,u_j^-\rangle$, $\langle u_i^+,b_j^-\rangle$ or $\langle b_i^+,b_k^-\rangle$, $i\ne k$. This is illustrated in Fig. \ref{fig:graph}. We call a circuit, say $C$, a path from the output to the input of a user, i.e. $C=\langle u^+,u^-\rangle$. In Fig. \ref{fig:graph}, the only possible circuit is given by $C=\langle u^+,b^-,b^+,u^-\rangle$.

\subsection{Fluid-Flow Paradigm}\label{sec:ex0}

In this paper, we shall present several laws allowing for the congestion modeling in a fluid-flow framework \cite{Lindley:52,Vinnicombe:00b,Misra:00,Hollot:01}. Using such a groundwork, the modeling problem admits a scalable solution in which the network asynchrony is captured through appropriate expansions and compressions of the time-space. It also enables the use of well-established mathematical tools: real functions analysis, integration theory, dynamical systems, difference and differential equations, etc. A first conclusion is that a proper definition for flows of data can easily be stated.
\begin{define}
  Given any edge $E$ of the network, a nonnegative integrable scalar flow $\phi(x,s)$, $(x,s)\in E\times\mathbb{R}_+$ and time instants $t_0,t\in\mathbb{R}_+$, $t\ge t_0$, we have
\begin{equation}\label{eq:Nphi}
  N_x(t,t_0)=\int_{t_0}^t\phi(x,s)ds
\end{equation}
where $N_x(t,t_0)$ is a packet counter, i.e. the number of packets that have passed through point $x$ between $t_0$ and $t$ is given by $N_x(t,t_0)$, and the integral is a standard one, e.g. the Lebesgue integral.
\end{define}

Unlike electrical circuits where one flow (the current) circulates on the edges, a vector-valued flow circulates on the edges of the graph representation of the network. Using the notation defined in Section \ref{sec:netgraph}, we can build the flow vectors $\phi(x,t)$ using the '$\col$' operator\footnote{which stacks in column the arguments.}:
\begin{equation}\label{eq:flow-+}
\begin{array}{rcl}
    \phi(x,t)&=&\col_{k=1}^{\sigma(x)}\left[\phi_k(x,t)\right],\\
    x&\in&\bigcup_i\bigcup_j\{u_i^-,u_i^+\}\cup\{b_j^-,b_j^+\}
\end{array}
\end{equation}
where $\sigma(x)$ is the number of flows in parallel using node $x$. Note that the nodes then implement multiplexers and demultiplexers for appropriate routing of the flows.

\subsection{Law 1: Information Conservation Law}\label{sec:ax1}

The first law is essentially a conservation law relating flow integration on two different domains. This conservation law follows from the remark that the quantity of information is preserved in a communication network: the data is either in transit, lost or received. Assuming lossless networks, it is possible to determine the total number of packets in transit on any edge, simply by counting the number of entering packets over a certain time horizon.

\begin{law}\label{ax:1}
\emph{Given any edge $E$ of the network, then for all $t\in\mathbb{R}_+$ there exists a time $t_0(t)\in\mathbb{R}_+$, $t_0(t)\le t$ such that
\begin{equation}\label{eq:lolmdr}
\begin{array}{lcl}
P_E(t)&:=&\int_{E}\phi(\theta,t)d\theta\\
&=&\int_{t_0(t)}^t\phi(\beta(E),s)ds\\
&=&N_{\beta(E)}(t,t_0(t))
\end{array}
\end{equation}
The integration over $E$ is an abstract integral which has to be understood as a flow integration from $\beta(E)$ to $\eps(E)$, that is, the number of packets $P_E(t)$ in the edge $E=\langle\beta(E),\eps(E)\rangle$ at time $t$.\mendlaw}
\end{law}

The main features of this law are the domain of integration exchange and the discretization of the spatial domain to nodes only. These considerations dramatically simplifies the modeling since it is no longer necessary to consider the flows at any point $x\in E$ but only at input nodes $\beta(E)$. This is illustrated by the following proposition:
\begin{proposition}\label{prop:ax1}
   The input flow $\phi(\beta(E),\cdot)$ and the output flow $\phi(\eps(E),\cdot)$ of edge $E$ verify
   \begin{equation}
  \phi(\eps(E),t)=t_0(t)^\prime\phi(\beta(E),t_0(t)).
\end{equation}
\end{proposition}
\begin{proof}
Since $N_{\beta(E)}(t,t_0(t))$ is the current number of packets in the edge $E$ at time $t$, then differentiation with respect to time provides the corresponding rate of variation
  \begin{equation*}
  [N_{\beta(E)}(t,t_0(t))]^\prime=\phi(\beta(E),t)-t_0(t)^\prime\phi(\beta(E),t_0(t)).
\end{equation*}
Moreover, the variation of the number of packets verifies
  \begin{equation*}
  [N_{\beta(E)}(t,t_0(t))]^\prime=\phi(\beta(E),t)-\phi(\eps(E),t)
\end{equation*}
which is nothing else but the difference between the input and output flows. The result follows from identification of the equalities.
\end{proof}

This proposition turns out to be very useful to derive models for transmission channels and buffers.
%The term $\bar{\rho}(t)$ is here to account for possible data losses during propagation. Indeed, while the reliability profile $\rho(x,t)$ acts in a distributed way along the link, the input profile acts locally at the input only. Roughly speaking, rather than losing bit by bit all along the edge, we can say that we lose once and for all, at the entrance, what will be lost during the propagation.

%It is difficult to provide a general expression for the function $\bar{\rho}$ depending on $\rho(x,t)$. However, we will see that in most of the interesting cases, such a function can be easily computed.

\subsection{Law 2: Queues are Flow Integrators}

The law given below defines the behavior of queues involved, for instance, inside routers and servers. Following past works and our understanding of the problem, the integrator model for queues is the most realistic.
\begin{law}\label{ax:2}
\emph{The queue dynamics of buffer $i$ is governed by the model
\begin{equation}\label{eq:buffer}
  \dot{q}_i(t)=\sum_j\phi_j(b_i^-,t)-r_i(t)
\end{equation}
with aggregated output flow rate
\begin{equation}\label{eq:outrate}
    r_i(t)=\left\{\begin{array}{lcl}
      c_i & &\mathrm{if}\ \mathcal{C}_i(t)\\
      \sum_j\phi_j(b_i^-,t) & & \mathrm{otherwise}.
    \end{array}\right.
\end{equation}
Above, $q_i$, $c_i$ and $\phi_j(b_i^-,t)$ represent the queue size, the maximal output capacity and the flow of type $j$ at the input, respectively. The condition $\mathcal{C}_i(t)$ is given by
\begin{equation}
  \mathcal{C}_i(t):=\left(\left[q_i(t)>0\right]\vee\left[\sum_j\phi_j(b_i^-,t)>c_i\right]\right).
\end{equation}
The corresponding queuing delay can be easily deduced using the relation $\tau_i(t)=q_i(t)/c_i$.}
\mendlaw
\end{law}
The above model can be proved to be the small packet limit of an M/M/1 queue \cite{Moller:08}. It can also be refined to capture additional features like finite maximal queue length, flow priorities, multiple output capacities, etc.

\subsection{Law 3: Users Model Existence}

The last law concerns the user protocol description and the way it dynamically reacts to congestion in the network.

\begin{law}\label{ax:3}
\emph{There exist bounded functions $\mathcal{P}_i$, $\mathcal{W}_i$ and $\mathcal{U}_i$ such that the trajectories $(z_i(t),w_i(t))$ of the following continuous-time model defined over $t\in\mathbb{R}_+$
\begin{equation}\label{eq:protocol}
\begin{array}{lcl}
    \dot{z}_i(t)&=&\mathcal{P}_i(z_i(t),\mu_i(t))\\
    w_i(t)&=&\mathcal{W}_i(z_i(t),\mu_i(t))\\
    \phi_i(u_i^+,t)&=&\mathcal{U}_i(w_i(t),\phi_i(u_i^-,t))
\end{array}
\end{equation}
match the trajectories of the asynchronous protocol\footnote{The asynchronous discrete decision instants of the actual protocol are assumed to belong to a countable set $\mathbb{T}_i$.} at points in $\mathbb{R}_+\cap\mathbb{T}_i$. Above, $z_i$, $\mu_i$, $\phi_i(u_i^-,\cdot)$ and $\phi_i(u_i^+,\cdot)$ are the state of the protocol, the measurements, the acknowledgment flow rate and the user sending flow respectively. The congestion window size $w_i$ is considered here as the number of outstanding packets to track and is supposed to be (weakly) differentiable.}
\mendlaw
\end{law}

%It may seem strange to define the protocol as a continuous-time dynamical system at first sight. Actually, as emphasized in the universal-clock discussion, the framework we will develop further on relies on a continuous-time modeling.
A procedure to solve the above interpolation problem has been first proposed in \cite{Misra:00} for TCP and reused in \cite[Appendix C.]{Jacobsson:09} for FAST-TCP. 

\section{Preliminary results}\label{sec:immediate}
\subsection{Transmission Channel With Constant Propagation Delay}\label{sec:transmed}

%\subsection{Input profiles}
%
%We will assume here that the signals propagate at constant speed $v$ in a transmission medium $M$ of effective length $\ell$ and reliability profile $\rho(x,t)$. The starting and ending nodes are denoted by $\beta(M)$ and $\eps(M)$ respectively and the propagation delay is given by $T=\ell/v$. In such a case, we have the following rule of conversion to obtain the input profile $\bar{\rho}$ from the reliability profile $\rho$:
%\begin{equation}
%  \bar{\rho}(t)=\frac{1}{T}\int_{\beta(M)}^{\eps(M)}\rho(x,t-v^{-1}x)dx
%\end{equation}

%\subsection{An operator for transmission media with constant propagation delay}

The following result concerning transmission channels is an immediate consequence of law \ref{ax:1}:
\begin{result}
Given a lossless transmission channel, corresponding to an edge $E$, with constant propagation delay $T>0$, the output flow is given by
\begin{equation}
  \phi(\eps(E),t)=\phi(\beta(E),t-T).
\end{equation}
%\mendprop
%The output flow hence consists of a delayed-version of the input flow.
\end{result}
\begin{proof}
Following law \ref{ax:1}, the number of packets in transit $P_E(t)$ in the edge $E$ at time $t\in\mathbb{R}_+$ obeys
\begin{equation}
  \begin{array}{lcl}
    P_E(t)&=&\int_E\phi(x,t)dx\\
    &=&\int_{t_0(t)}^t\phi(\beta(E),s)ds\\
    &=&N_{\beta(E)}(t,t_0(t)),
  \end{array}
\end{equation}
where $t_0(t)=t-T$ since the propagation delay is constant. Indeed, a packet sent at time $t-T$ will, at time $t$, still be on the edge but about to leave. The result follows then from Proposition \ref{prop:ax1}.
%
% Differentiating the above expression, we get
%\begin{equation}
%\begin{array}{c}
%  N_E(t)^\prime=\phi(\beta(E),t)-\phi(\beta(E),t-T).
%\end{array}
%\end{equation}
%Furthermore, the variation of the number of packets in the medium also obeys the following rule
%\begin{equation}
%  N_E(t)^\prime=\phi(\beta(E),t)-\phi(\eps(E),t)
%\end{equation}
%which is nothing else but the difference between the input and output flows; see Section \ref{sec:ax1}. Identifying the two equalities yields the result.
\end{proof}

%Lossy channels can also be characterized by introducing a flow input-output gain $\bar{\rho}_E(t)$ as
%\begin{define}
%  The proportional loss factor $\bar{\rho}(\cdot)\in[0,1]$ of a transmission medium, corresponding to an edge $E$ with constant propagation delay $T>0$, is defined as the ratio
%\begin{equation}
%  \bar{\rho}_E(t):=\frac{\phi(\eps(E),t)}{\phi(\beta(E),t-T)}.
%\end{equation}
%\end{define}
%

\begin{figure}
  \centering
        \psfrag{d}[c][c]     {\small{Constant propagation delay $T$}}
        \psfrag{ch}[c][c]     {\small{\shortstack[l]{Transmission channel\\$\qquad\quad \langle u^+,b^-\rangle$}}}
        \psfrag{ip}[c][c]     {\small{$\phi(u^+,t)$}}
        \psfrag{op}[c][c]     {\small{\shortstack{$\phi(b^-,t)$\\$=$\\$\phi(u^+,t-T)$}}}
  \includegraphics[width=0.35\textwidth]{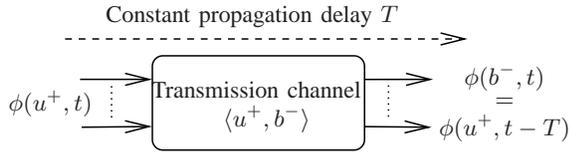}
  \caption{Transmission channel block}\label{fig:channel}
\end{figure}

\subsection{The ACK-Clocking model}\label{sec:ack}

The ACK-clocking model \cite{Jacobsson:08,Jacobsson:08b,Tang:10} is certainly the most important consequence of law \ref{ax:1}. This model characterizes the \emph{flight-size}\footnote{The number of outstanding packets.} $\digamma_i(C_i,t):=P_{C_i}(t)$ of a user $i$ at any time $t\in\mathbb{R}_+$ over a closed circuit $C_i=\langle u_i^+,u_i^-\rangle$. The importance of the ACK-clocking model lies in the semantic it adds to the model by relating RTT, flow and flight-size together\footnote{Note however than in \cite{Tang:10} the congestion window size is considered instead of the flight-size, which is rather different. Equivalence holds when some conditions, such as the condition flight-size$\ge$congestion window size, are met. We will come back on this in Section \ref{sec:userwinflight}.}. From this result, it will be possible to derive a number of important properties and rules for the users.

\begin{result}[ACK-Clocking]
The ACK clocking model is given by
\begin{equation}
\begin{array}{lcl}
      \digamma_i(C_i,t+\RTT_i\{t\}) &=& \displaystyle{\int_{C_i}\phi(\theta,t+\RTT_i\{t\})d\theta}\\
                        &=& \smashoperator{\int_{t}^{t+\RTT_i\{t\}}}\phi_i(u_i^+,s)ds
\end{array}
\end{equation}
where $\RTT_i\{t\}$ is the RTT of a packet sent at time $t$ in the circuit $C_i$ by user $i$.\menlem
\end{result}
\begin{proof}
  We assume here that the circuit is lossless for simplicity. The flight-size is indeed a spatial integration of flows since the number of packets in transit is equal to the spatial integral of the flows over the circuit. Thus, following law \ref{ax:1}, it is possible to convert the spatial integration into a temporal one provided that we can determine the integration domain. To obtain it, we use the notion of RTT and suppose that a data sent by user $i$ in the circuit $C_i$ at time $t$ has a round-trip-time given by $\RTT_i\{t\}$. This means that the data sent between $t$ and $t + \RTT_i\{t\}$ are still unacknowledged at $t+\RTT_i\{t\}$, thus still in the circuit. Hence, the corresponding temporal integration has bounds $t$ and $t+\RTT_i\{t\}$.
\end{proof} 

\section{RTT expression and internal FIFO buffer model}\label{sec:buffer}
In the light of the discussion above, it turns out that a model for the RTT is necessary in order to characterize and compute the flight size $\digamma_i(C_i,t+\RTT_i\{t\})$. A first step forward towards a RTT expression is the analysis of queuing delays in buffers.

\subsection{Forward/Backward Delays and Causal RTT Expression}\label{sec:delayRTT}

Since RTT directly depends on the queuing delays, its computation essentially relies on calculating the queueing delays. The results are taken from a previous work of us \cite{Briat:10} with the difference that we relate them here to law \ref{ax:2}.

Considering the buffer model (\ref{eq:buffer}) with queueing delay $\tau_i(t)$, we define the \emph{forward delay operator} $f_i(t):=t+\tau_i(t)$ which maps, at a flow level, any packet input time $t$ to the corresponding packet output time $f_i(t)$. Although it is easy to derive and understand, this operator leads to a noncausal RTT expression which is not desirable. To observe this, let us consider a closed circuit $C$ with $N$ queues, indexed from 1 to $N$. The indices $0$ and $N+1$ are used to denote the input and output of the circuit respectively. Given a packet input time $t$, the corresponding packet output time $t_C$ is given by:
\begin{equation}
\begin{array}{lcl}
    t_C(t)&=&\mathscr{F}_C(t)\\
    \mathscr{F}_C&=& \mathcal{R}^{-1}_{N,N+1}\circ f_{N}\circ\mathcal{R}^{-1}_{N-1,N}\circ f_{N-1}\circ \ldots\\
    &&\circ\mathcal{R}^{-1}_{2,3}\circ f_{2}\circ\mathcal{R}^{-1}_{1,2}\circ f_{1}\circ\mathcal{R}^{-1}_{0,1}\circ ev
\end{array}
\end{equation}
where $\mathcal{R}_{i,j}$ corresponds to the transmission channel operator (constant delay operator) between queues $i$ and $j$, $ev$ is the evaluation map and $\circ$ is the composition operator. The same formula, albeit expressed in different ways, has been also obtained in \cite[Section 3.3.5]{Jacobsson:08} and \cite[equations (7d-7f)]{Tang:10}.

It is clear that  $\mathscr{F}_C$ is noncausal since it requires the knowledge of future information, which is not available. In order to solve this problem, the \emph{backward delay operator} \cite{Briat:10} expressing the input time $t$ as a function of the output time $t_C$, is considered instead. Hence, the problem reduces to inverting the forward delay operator. The existence of this inverse operator and some of its properties are recalled below:
\begin{result}[\cite{Briat:10}]
  The operator $f_i$ is invertible if and only if the input flow of the corresponding buffer is almost everywhere positive.\menlem
\end{result}
\begin{result}[\cite{Briat:10}]\label{lem:fg}
The functions $g_i:=f_i^{-1}$ obeys:
    \begin{subequations}\label{eq:fg}
   \begin{eqnarray*}
%    f_i(t)&=&t+\tau_i(t)\label{eq:prop1}\\
    g_i(t)&=&t-\tau_i(g_i(t))\label{eq:prop2}\\
    g_i^\prime(t)&=&\left\{\begin{array}{ll}
    \displaystyle{c_i\left(\sum_{k=1}^{\sigma(b_i^-)}\phi_k(b_i^-,g_i(t))\right)^{-1}} &\mathrm{if\ }\mathcal{C}_i(g_i(t))\\
    1 &\mathrm{otherwise}
  \end{array}\right.\label{eq:prop3}
   % f_i^\prime(t)&=&\left\{\begin{array}{lcl}
%    \displaystyle{c_i^{-1}\sum_{k=1}^{\sigma(b_i^-)}\phi_k(b_i^-,t)} &&\mathrm{if\ }\mathcal{C}_i(t)\\
%    1 &&\mathrm{otherwise}
%  \end{array}\right.\label{eq:prop4}%\\
 % {[\tau_i(g_i(t))]}^\prime&=&\left\{\begin{array}{lcl}
%    1-c_i\left(\sum_{k=1}^{\sigma(b_i^-)}\phi_k(b_i^-,g_i(t))\right)^{-1}&\mathrm{if\ }\mathcal{C}_i(g_i(t))\\
%    1&&\mathrm{otherwise}
%  \end{array}\right.\label{eq:prop5}
  \end{eqnarray*}
\end{subequations}
where $g_i^\prime(t)$ stands for the the upper right Dini derivative of $g_i(t)$, i.e. $D^+[g_i](t)=\limsup_{h\downarrow0}h^{-1}\left(g_i(t+h)-g_i(t)\right).$\menlem
\end{result}

Using the backward delay operators $g_i$, the packet sending time $t$ can be computed from the reception time $t_C$ through the causal expression:
\begin{equation}
\begin{array}{cll}
    t(t_C)&=&\mathscr{B}_C(t_C)\\
%    &=&t_C-T_C-\tau_C\left(\mathscr{B}_C(t_C)\right)\\
    \mathscr{B}_C&=&\mathcal{R}_{0,1}\circ g_{1}\circ\mathcal{R}_{1,2}\circ g_{2}\circ\mathcal{R}_{2,3}\circ\ldots\\
    &&\circ g_{N-1}\circ\mathcal{R}_{N-1,N}\circ g_{N}\circ\mathcal{R}_{N,N+1}\circ ev.
\end{array}
\end{equation}

\begin{example}
  In the single-user/single-buffer case, the expressions reduce to
  \begin{equation}
  \begin{array}{lcl}
        t(t_C)&=&g(t_C-T_b)-T_f\\
        t_C&=&t+T_f+T_b+\tau(t+T_f)\\
        \RTT\{t\}&=&t_C-t\\
        &=&t_C-g(t_C-T_b)+T_f\\
        &=&T_b+T_f+\tau(g(t_C-T_b))
  \end{array}
  \end{equation}
  where $T_f$ and $T_b$ are the forward and backward propagation delays corresponding to $\mathcal{R}_{0,1}$ and $\mathcal{R}_{1,2}$.
\end{example}

Using the backward expression of the RTT, it easy to obtain the following result:
\begin{result}
%\begin{myprop}
  The flight size obeys
  \begin{equation}\label{eq:FS}
    \begin{array}{rcl}
      \digamma_i(C_i,\mathscr{F}_{C_i}(t)) &=& \smashoperator{\int_{t}^{\mathscr{F}_{C_i}(t)}}\phi(u_i^+,s)ds\\
      \digamma_i(C_i,t) &=& \smashoperator{\int_{\mathscr{B}_{C_i}(t)}^{t}}\phi(u_i^+,s)ds.\qquad\qquad\quad\mendprop
    \end{array}
  \end{equation}
\end{result}
%\end{myprop}

This is a direct consequence of law \ref{ax:1} (through the ACK-clocking model) and law \ref{ax:2}. However, the problem is not completely resolved yet since it is still rather unclear how the queuing delays along a given circuit may be computed. Indeed, calculating the queueing delays requires the knowledge of the input flows of all buffers, and hence necessitates a way of splitting the upstream buffer aggregated output flows into distinct 'atomic' flows. Otherwise, a modular description of buffer interconnections is not possible.

\subsection{FIFO Buffer Output Flow Separation}\label{sec:outputflow}

Without further considerations on the queue type, there exists an infinite number of ways to separate the aggregated output flow directly from the queuing model of law \ref{ax:2}. When a FIFO queue (i.e. order preserving) is considered, it turns out that the output flow separation problem is easily solvable using laws \ref{ax:1} and \ref{ax:2}. The FIFO characterization and output flow separation problems have been fully solved in \cite{Briat:10}. In this section, we will recall and explain these results and connect them to laws \ref{ax:1} and \ref{ax:2}.
\begin{result}[\cite{Briat:10}]
  Let us consider the queueing model (\ref{eq:buffer}) which we assume to represent a FIFO queue. The output flow corresponding to the input flow $\phi_\ell(b_i^-,t)$ is given by
  \begin{equation}\label{eq:bufferoutput}
  \begin{array}{lcl}
    \phi_\ell(b_i^+,t)&=&g_i^\prime(t)\phi_\ell(b_i^-,g_i(t))\\
    &=&\left\{\begin{array}{lcl}
      \dfrac{c_i\phi_\ell(b_i^-,g_i(t))}{\sum_j^{\sigma(b_i^-)}\phi_j(b_i^-,g_i(t))}&&\mathrm{if\ }\mathcal{C}_i(t)\\
      \phi_\ell(b_i^-,t)&&\mathrm{otherwise}
    \end{array}\right.
  \end{array}
  \end{equation}
\end{result}
\begin{proof}
  The proof is available in \cite{Briat:10} and is based on the analysis of the contribution of each input flow to the queue size. Then, using laws \ref{ax:1} and \ref{ax:2}, it is possible to split the aggregate output flow into atomic output flows that correspond to each input flow.
\end{proof}

This model\footnote{Such a model has also been proposed in \cite{Liu:04, Ohta:98} without any proof. It is proved here that this model is an immediate consequence of the conservation law \ref{ax:1}.} deserves interpretation: the output flows consist of scaling and shifting of the input flows. The delay accounts for the high flow viscosity and captures the queue FIFO behavior, \emph{at the flow level}. This model also tells that the output flow corresponding to the input flow $\phi_\ell(b^-,t)$ is expressed as a (delayed) ratio of the input flow $\phi_\ell(b^-,t)$ to the total input flow that entered the buffer at the same time. Hence, the output flows are proportional to relative flows modeling the probability of having a packet of certain type served at time $t$. This probability is then scaled-up by the maximal output capacity to utilize the available bandwidth. This nonlinear expression for the output flows describes the flow-coupling, through the nonlinear expression in (\ref{eq:bufferoutput}), and clock-coupling phenomena discussed in Section \ref{sec:ex0}, through the delay dynamical model depending on input flows. Note also that the complete model (\ref{eq:buffer})-(\ref{eq:bufferoutput}) is a quite unusual dynamical system consisting of a delayed direct feedthrough operator whose delay is implicitly defined by a dynamical expression.

\begin{figure}
  \centering
        \psfrag{d}[c][c]     {\small{Forward queueing delay $\tau$}}
        \psfrag{b}[c][c]     {\small{Backward queueing delay $\tau(g)$}}
        \psfrag{ch}[c][c]     {\small{Queue $\langle b^-,b^+\rangle$}}
        \psfrag{ip}[c][c]     {\small{$\phi(b^-,t)$}}
        \psfrag{op}[c][c]     {\small{\shortstack{$\phi(b^+,t)$\\defined in (\ref{eq:bufferoutput})}}}
  \includegraphics[width=0.35\textwidth]{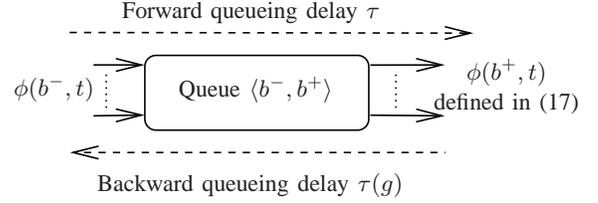}
  \caption{Queue/Buffer block}\label{fig:buffer}
\end{figure}

\begin{example}
To illustrate the model, let us consider two input flows given by $\phi_1(t)=0.55c\left(1+\Sq(t)\right)$ and $\phi_2(t)=0.55c\left(1-\Sq(t)\right)$%
%\begin{equation}
%  \begin{array}{lcl}
%    \phi_1(t)&=&0.55c\left(1+\Sq(t)\right),\\
%    \phi_2(t)&=&0.55c\left(1-\Sq(t)\right)
%  \end{array}
%\end{equation}
where $c=100$Mb/s and the function $\Sq(t)$ is a square function of period $T=1$s. Since the flows are in phase opposition and their sum constant, they lead to an alternation of packet types in the queue while respective packet populations remain very close to each other at any time. Therefore, the output flows should reflect the contents of the queues and the model should be able to keep track of the order of arrival of packets in the queue. The simulation results are plotted in Fig. \ref{fig:compflow} where we can see that the proposed model reflects exactly the contents of the queue. Indeed, the output flows are deformations of the input flows: they are scaled in amplitude and time, so that the surfaces below the curves (the number of packets) over one period are the same. This illustrates the law of conservation of the information. In Fig. \ref{fig:compflow2}, a comparison of the results obtained by the model (\ref{eq:buffer})-(\ref{eq:bufferoutput}) and NS-2 is shown. To overcome the problem of flow computation in NS-2, the number of input and output packets of the queue is computed for both users and compared. We can see that the model matches perfectly NS-2 simulations showing that the model well captures the internal organization of packets inside the queue. This also means that the output flow model is relevant and represents what really happens at the output of the queue.

\begin{figure}[h!]
\begin{minipage}[b]{0.49\linewidth}
\centering
 \includegraphics[width=\textwidth]{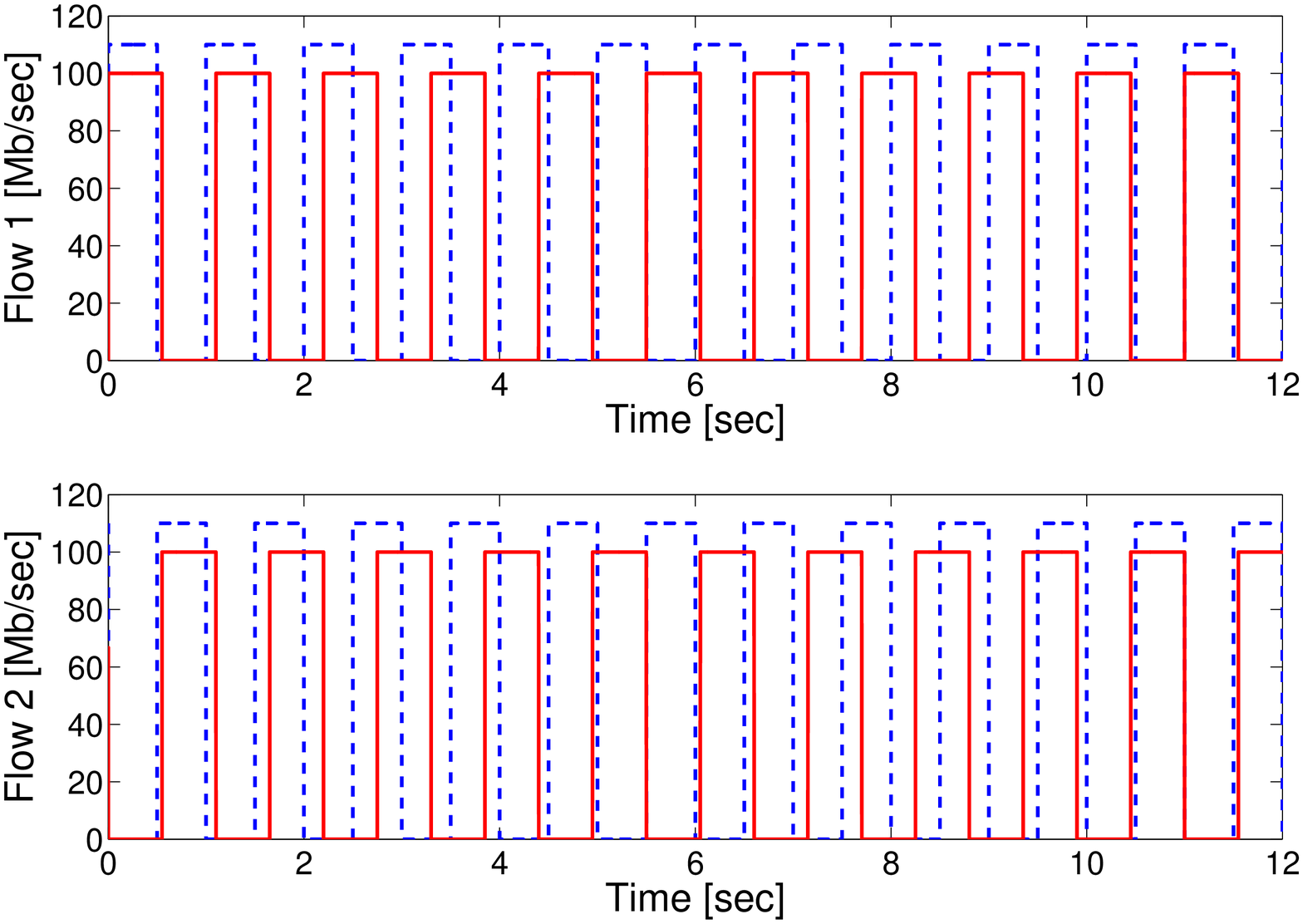}
  \caption{Predicted output flows. Plain: output flows, dashed: input flows}\label{fig:compflow}
\end{minipage}
\hfill
\begin{minipage}[b]{0.49\linewidth}
\centering
 \includegraphics[width=\textwidth]{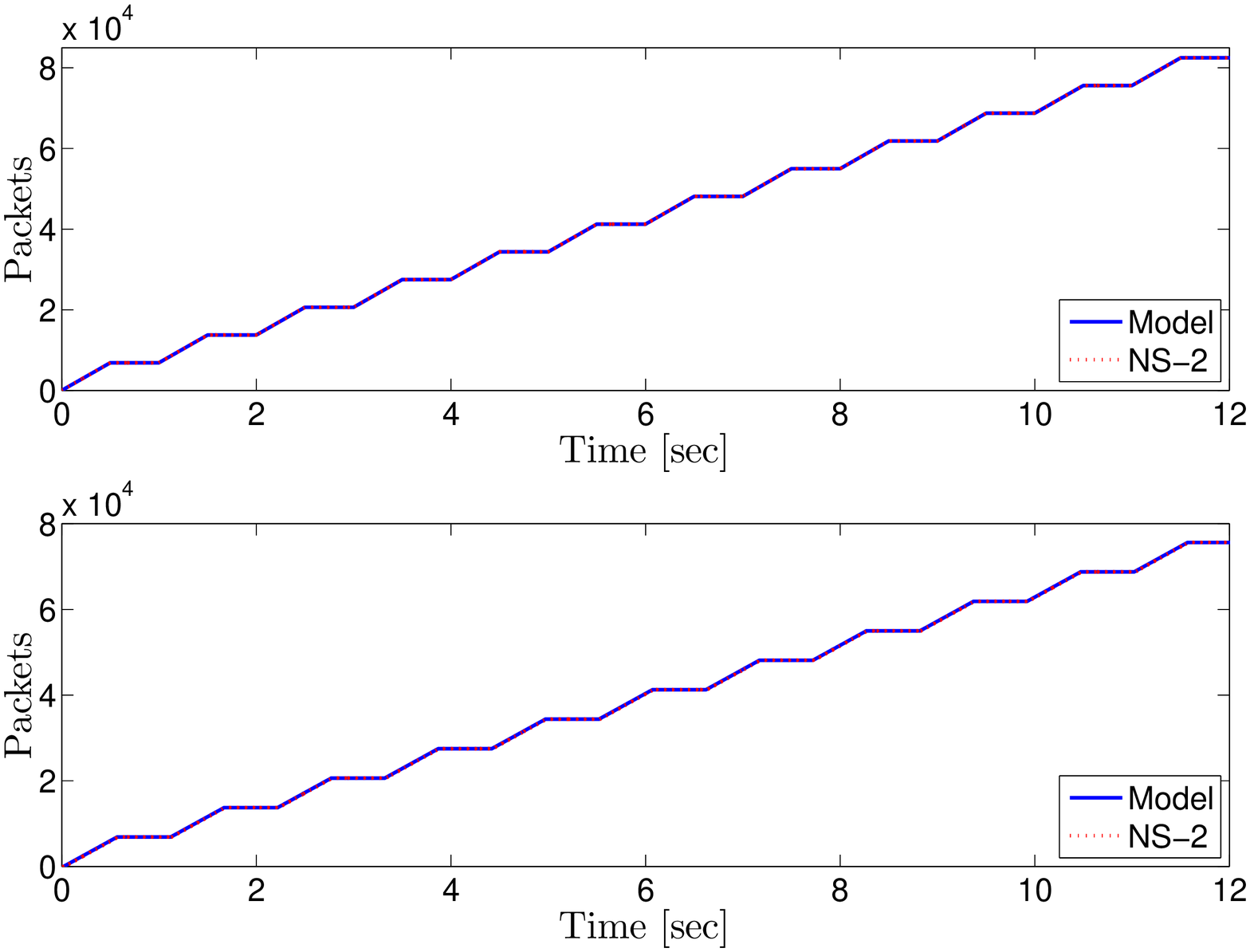}
\caption{Model predicted and NS-2 simulated input and output number of packets: flow 1 (top) and flow 2 (bottom).}\label{fig:compflow2}
\end{minipage}
\end{figure}
%
%
%The model (\ref{eq:modelq}) is valid when the flows are slowly varying, e.g. around equilibrium or for sufficiently long piecewise constant flows. As illustrated in Fig. \ref{fig:compflow}, it is then not well-suited for an accurate description of queues in the transient phase. The queueing delay is not initially integrated in the modeling and should be incorporated afterwards, leading then to complications since telling which delay value affects output flow is difficult due to the non-capture of the packet order of arrival. Finally, the need of keeping track of all the pseudo-queues, one for each input flow, may be a source of complexity.
\end{example}

\section{A New user model}\label{sec:user}
As explained in the introduction, the user modeling problem is partially an open question and a solution, based on Laws \ref{ax:1} and \ref{ax:3}, is proposed here. The distinct notions of flight-size, ACK-flow and congestion window size are clarified first and associated with each other. Then, the problem of computing users sending flow $\phi(u_i^+,t)$ is solved. Finally, the congestion-window-to-flight-size conversion problem, accounting for flight-size rate of variation constraints, is addressed.

\begin{figure}[H]
  \centering
        \psfrag{ch}[c][c]     {\small{User $\langle u^-,u^+\rangle$}}
        \psfrag{m}[c][c]     {\small{$\mu(t)$}}
        \psfrag{ip}[c][c]     {\small{\shortstack{$\phi(u^-,t)$\\ defined in (\ref{eq:ackflow})}}}
        \psfrag{op}[c][c]     {\small{\shortstack{$\phi(u^+,t)$\\ defined in (\ref{eq:userflow1})}}}
  \includegraphics[width=0.35\textwidth]{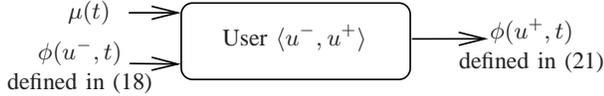}
  \caption{User block}\label{fig:user}
\end{figure}

\subsection{ACK-Clocking Dynamics and User Flow Computation}

The approach exposed here is based on the ACK-clocking model of Section \ref{sec:ack} obtained from Laws \ref{ax:1} and \ref{ax:2}.
\begin{result}
Let us consider a circuit $C_i=\langle u_i^+,u_i^-\rangle$. Then the ACK-flow the user $u_i$ receives is given by
  \begin{equation}\label{eq:ackflow}
    \phi(u_i^-,t)=\mathscr{B}^\prime_{C_i}(t)\phi_i(u_i^+,\mathscr{B}_{C_i}(t)).
  \end{equation}
  \mendprop
\end{result}
\begin{proof}
The key idea is to remark that $\digamma_i(C_i,t)=N_{u_i^+}(t,\mathscr{B}_{C_i}(t))$. Hence, using Proposition \ref{prop:ax1} and noting that the ACK-flow is the leaving flow from the circuit, we obtain the result.
%
%, we have the result.
%
%The key idea is to compute the time-derivative of the flight-size $\digamma_i(C_i,t)$ (\ref{eq:FS}) to get
%\begin{equation}\label{eq:dsqdsq}
%\digamma_i(C_i,t)^\prime=\phi_i(u_i^+,t)-\mathscr{B}_{C_i}(t)^\prime\phi_i(u_i^+,\mathscr{B}_{C_i}(t)),
%\end{equation}
%where the second term of the RHS is very similar to the output flow model for buffers (\ref{eq:bufferoutput}), at the difference that the derivative of the circuit backward operator is involved. Note also that, by virtue of law \ref{ax:1}, the flight-size derivative also obeys
%\begin{equation}
%  \digamma_i(C_i,t)^\prime=\phi_i(u_i^+,t)-\phi_i(u_i^-,t)
%\end{equation}
%where $\phi_i(u_i^-,t)$ is the circuit leaving flow, hence corresponding to the ACK-flow the user receives. The result follows then from identification of the two equalities.
\end{proof}
Note that the differentiation of (\ref{eq:FS}) also yields
\begin{equation}\label{eq:mdrlol1}
  \phi_i(u_i^+,t)=\digamma_i^\prime(C_i,t)+\mathscr{B}^\prime_{C_i}(t)\phi_i(u_i^+,\mathscr{B}_{C_i}(t))
\end{equation}
meaning that to maintain the same flight size the user has to naturally send data at the same rate as receiving ACK packets: \emph{this is exactly ACK-clocking but expressed at the flow level}. By flow integration, we can easily recover the 'packet-level ACK-clocking'. A similar formula, albeit not derived from any conservation law, is considered in \cite{Zhang:10}. It however uses a different model than (\ref{eq:bufferoutput}) for representing the buffer output flows\footnote{It can be shown that the output flow model of \cite{Zhang:10} does not match NS-2 simulations in general, but only in some specific cases, such as the constant flows case.} and involves the congestion window size instead of the flight-size.

\subsection{User Flow, Flight-Size and Congestion Window Size}\label{sec:userwinflight}

We clarify here the relation between the user sending flow $\phi(u_i^+,t)$ and the congestion window size $w_i(t)$. First, recall that the congestion window size corresponds to the desired flight-size while the flight-size is the current number of packets in transit. The congestion window size is then a \emph{reference} to track while the flight size is the \emph{controlled output}. The \emph{control input} is the user sending flow for which constraints must be considered.

Indeed, when the congestion window size increases, the user can immediately send a burst of new packets to equalize the flight-size and congestion window size. In such a case, we can ideally assimilate them to be equal (and so are their derivatives). The small delay corresponding to the protocol reaction time can be easily incorporated in the constant part of the RTT. The problem is, however, slightly more difficult when the congestion window size decreases and becomes smaller than the flight-size. In such a case, we cannot withdraw packets from the network and the only thing we can do is to wait for new ACK packets until the flight size becomes equal to the congestion window size. Therefore, while the slope of the flight-size is unbounded from above, it is basically bounded from below. In \cite{Jacobsson:08}, a rate-limiter is used to control the slope of the flight size but this is rather limited due to the absence of any ACK-flow model and the difficulty arising from the time-varying nature of the lower bound value to consider. In most recent works, such as \cite{Zhang:10,Tang:10}, this problem is excluded by considering that the flight-size is always smaller than the congestion window size and that the congestion window size does not decrease 'too rapidly'. We provide here an explicit solution to this problem based on a hybrid modeling of the user behavior.

According to the above discussion, the flight-size must obey
\begin{equation}\label{eq:flightsizecontrolled}
    \digamma_i(C_i,t)^\prime=\left\{\begin{array}{lcl}
      \dot{w}_i(t) && \mathrm{if\ }\mathcal{T}_i(t)\\
      -\phi(u_i^-,t)&& \mathrm{otherwise}
    \end{array}\right.
\end{equation}
where $\mathcal{T}_i(t)$ is a condition which is false when the flight-size must be decreased and true otherwise.

%We then have the following result:
\begin{result}
The flight-size $\digamma_i(C_i,t)$ satisfies (\ref{eq:flightsizecontrolled}) if the user sending rate is defined as
  \begin{equation}\label{eq:userflow1}
    \phi(u_i^+,t)=\left\{\begin{array}{lcl}
      \dot{w}_i(t)+\phi(u_i^-,t) && \mathrm{if\ }\mathcal{T}_i(t)\\
      0&& \mathrm{otherwise}
    \end{array}\right.
  \end{equation}
  where $\mathcal{T}_i(t)=\left([\pi_i(t)=0]\wedge[\dot{w}_i(t)+\phi(u_i^-,t)\ge0]\right)$ and
  \begin{equation}\label{eq:userflow2}
    \dot{\pi}_i(t)=\left\{\begin{array}{lcl}
      0 && \mathrm{if\ }\mathcal{T}_i(t)\\
      \dot{w}_i(t)+\phi(u_i^-,t) && \mathrm{otherwise}.
    \end{array}\right.
  \end{equation}
  Moreover, this model is the simplest one.\menlem
\end{result}
\begin{proof}
  The virtual buffer $\pi_i$, taking nonpositive values, measures the number of ACK packets to retain in order to balance the flight-size and congestion window size. When the virtual buffer has negative state, i.e. $\pi_i(t)<0$, the arriving ACK-packets have to be absorbed until the state reaches 0. Once zero is reached, the user can start sending again until the congestion window size decreases too rapidly, i.e. $\dot{w}_i(t)<-\phi(u_i^-,t)$. Substitution of the user sending rate defined by (\ref{eq:userflow1})-(\ref{eq:userflow2}) in (\ref{eq:mdrlol1}) yields the flight-size behavior (\ref{eq:flightsizecontrolled}). To see that the model is minimal, it is enough to remark that both conditions in $\mathcal{T}_i(t)$ are necessary.
\end{proof}

The state of the extended user model hence consists of both the state of the congestion controller $z_i$ and the ACK-buffer $\pi_i$.

%The user behavior also depends on the measurements $\mu_i(\varkappa_t)$, functions of the overall network state $\varkappa_t$; this state will be discussed in more details in Section \ref{sec:general}.
%
%We are now in a position to derive the users operators from the equations (\ref{eq:userflow1}).
%\begin{define}
%  The users operators $\mathcal{U}_i(w_i):\mathbb{R}_+\to\mathbb{R}_+$ mapping the ACK-flow $\phi(u_i^-,t)$ to the sending flow $\phi(u_i^+,t)$ is given by
%\begin{equation}
%  \phi(u^+,t)=
%   \mathcal{U}(w)%(\mu(\varkappa_t))
%  \phi(u^-,t)
%\end{equation}
%where $\mathcal{U}(w)=\diag_i\left\{\mathcal{U}_i(w_i)\right\}$ and $\mathcal{U}_i$ satisfies (\ref{eq:protocol}).
%\end{define} 

\section{Example of a single buffer topology}\label{sec:model}

\subsection{The Single-Buffer/Multiple-User Topology with FAST-TCP protocol}

In this section, we consider a single-buffer/multiple-users topology interconnected by lossless transmission channels. The forward and backward propagation delays of user $i$ are denoted by $T_i^f$ and $T_i^b$ respectively. For illustration, we use the following FAST-TCP model as user protocol model
\begin{equation}\label{eq:FAST}
    \dot{w}_i(t)=\gamma\left[-\dfrac{\tau(g^i(t))}{T_i+\tau(g^i(t))}w_i(t)+\alpha\right]
\end{equation}
where $w_i(t)$, $T_i=T_i^f+T_i^b$ and $g^i(t)=g(t-T_i^b)$ are the congestion window size, the total propagation delay and the backward queuing delay respectively.

\subsubsection{General Model}

The general model is given by (\ref{eq:FAST}) and

\begin{equation}\label{eq:general}
\small\begin{array}{lcl}
  \dot{\tau}(t)&=&\left\{\begin{array}{lcl}
    c^{-1}\eta(t)+\delta(t)-1&& \mathrm{if\ }\mathcal{C}(t)\\
    0 && \mathrm{otherwise}
  \end{array}\right.\\
  \dot{\pi}_i(t)&=&\left\{\begin{array}{lcl}
      0 && \mathrm{if\ }\mathcal{T}_i(t)\\
      \dot{w}_i(t)+\phi(u_i^-,t) && \mathrm{otherwise}.
    \end{array}\right.\\
  \digamma_i(t)&=&\int_{g^i(t)-T_i^f}^t\phi(u_i^+,\theta)d\theta\\
  \phi(u_i^+,t)&=&\left\{\begin{array}{lcl}
      \dot{w}_i(t)+\phi(u_i^-,t) && \mathrm{if\ }\mathcal{T}_i(t)\\
      0&& \mathrm{otherwise}
    \end{array}\right.\\
  \phi(u_i^-,t)&=&\left\{\begin{array}{lcl}
    \frac{c\phi(u_i^+,g^i(t)-T_i^f)}{c\delta(g^i(t))+\sum_j\phi(u_j^+,g^i(t)-T_j^f)}&& \mathrm{if}\ \mathcal{C}(t)\\
    \phi(u_i^+,t-T_i^b-T_i^f) &&\mathrm{otherwise}
  \end{array}\right.\\
   \eta(t)&=&\sum_{i=1}^{N}\phi(u_i^+,t-T_i^f)
\end{array}
\end{equation}
where $\delta(t)$ denotes the normalized cross-traffic $\delta(t)\in[0,1)$.
%The equilibrium point of this model is unique and is given by
%\begin{equation*}
%\begin{array}{lcllcl}
%    w_i^*&=&\alpha\left(1+\dfrac{T_i}{\tau^*}\right),&\tau^*&=&\dfrac{N\alpha}{c(1-\delta^*)}
%\end{array}
%\end{equation*}
%where $T_i=T_i^f+T_i^b$, $\delta^*\in[0,1)$ and $N$ are the equilibrium normalized cross-traffic, the propagation delay and the number of users respectively. Note that the equilibrium point of the network is both fair and efficient \cite{Sandberg:09}.
%In what follows, we will show that this model reduces into models of the literature when some particular conditions are met.

\subsubsection{Homogeneous Delays and No-Cross Traffic - The Static-Link Model}\label{sec:hnct}

In \cite{Tang:10}, it is shown that the ratio-link and the joint link models are actually approximations of the ACK-clocking model\footnote{involving the congestion window size instead of the flight-size.}, i.e. 0th and 1st order approximations of it. The static-link model \cite{Wang:05} is however retrieved after linearization and Pad\'{e} approximations of exponentials corresponding to delays. This 'proof' makes the static-link model only valid very locally. Below we show that the static link model is exact in some certain situations and we give a proof for its domain of validity.

In the case of homogeneous delays, i.e. $T_i^f=T^f$, $T_i^b=T^b$, $i=1,\ldots,N$, and absence of cross-traffic, i.e. $\delta\equiv0$, model (\ref{eq:general}) reduces to
\begin{equation}
\begin{array}{lcl}
  \dot{\tau}(t)&=&\left\{\begin{array}{lcl}
    c^{-1}\eta(t)-1&& \mathrm{if\ }\mathcal{C}(t)\\
    0 && \mathrm{otherwise}
  \end{array}\right.\\
\dot{\pi}_i(t)&=&\left\{\begin{array}{lcl}
      0 && \mathrm{if\ }\mathcal{T}_i(t)\\
      \dot{w}_i(t)+\phi(u_i^-,t) && \mathrm{otherwise}.
    \end{array}\right.\\
\digamma_i(t)&=&\int_{g_b(t)-T^f}^t\phi(u_i^+,\theta)d\theta\\
\phi(u_i^+,t)&=&\left\{\begin{array}{lcl}
      \dot{w}_i(t)+\phi(u_i^-,t) && \mathrm{if\ }\mathcal{T}_i(t)\\
      0&& \mathrm{otherwise}
    \end{array}\right.\\
     \phi(u_i^-,t)&=&\left\{\begin{array}{lcl}
    \frac{c\phi(u_i^+,g_b(t)-T^f)}{\sum_j\phi(u_j^+,g_b(t)-T^f)}&& \mathrm{if}\ \mathcal{C}(t)\\
    \phi(u_i^+,t-T^b-T^f) &&\mathrm{otherwise}
  \end{array}\right.\\
\eta(t)&=&\sum_{i=1}^{N}\phi(u_i^+,t-T^f)\\
      g_b(t)&=&g(t-T^b).
\end{array}
\end{equation}
Assuming the buffer is always congested (i.e. $\mathcal{C}(t)$ holds true) and all the users are active (i.e. the $\mathcal{T}_i(t)$'s are true) we obtain
%\begin{equation}
%\begin{array}{lcl}
%    \dot{\tau}(t)&=&c^{-1} \sum_i\sigma_i(t-T^f)\dot{w}_i(t-T^f)\\
%  &&+\bar{\sigma}_i(t-T^f)\phi(u_i^-,t-T^f)
%\end{array}
%\end{equation}
%where $\sigma_i(t)=1$ when $\mathcal{T}_i(t)$ is true, 0 otherwise and $\bar{\sigma}_i(t)=1-\sigma_i(t)$. When $\mathcal{T}_i(t)$ holds true for all users, we get
\begin{equation}
  \dot{\tau}(t)=c^{-1}\sum_i\dot{w}_i(t-T^f)
\end{equation}
which is the static-link model. Integrating the above equation from $0$ to $t$ we obtain
\begin{equation}\label{eq:sl}
  \tau(t)=c^{-1}\sum_iw_i(t-T^f)-T^f-T^b
\end{equation}
where we assumed $\tau(0)=0$ and $w_i(0)=0$, $i=1,\ldots,N$. Thus, according to the proposed model (\ref{eq:general}), the static-link model is valid whenever the necessary and sufficient conditions are met
\begin{itemize}
  \item the buffer is permanently congested, i.e. $\mathcal{C}(t)$ holds true;
  \item the propagation delays are homogeneous, i.e. $T_i^f=T^f$, $T_i^b=T^b$, $i=1,\ldots,N$;
  \item the cross-traffic is absent, i.e. $\delta\equiv0$;
  \item the users are not in ACK-retaining mode, i.e. $\mathcal{T}_i(t)$ holds true for all $i=1,\ldots,N$.
\end{itemize}

%It is interesting to note that these conditions are necessary and sufficient for the static-link model validity. The first one ensures that the queue is a bottleneck and is always congested. The second condition makes the sum of the ACK-flow contribution in the user sending flows equal to the queue output capacity, so that only the windows derivatives remain in the delay dynamical model. Note that in presence of heterogeneous delays, the sum may exceed the queue maximal output capacity. The absence of cross-traffic also ensures that only the windows derivatives remain. Finally, if the user is in ACK-retaining mode, then it does not send any flow and the window derivative term disappears from the model. In \cite{Tang:10} an attempt for justifying the static-link model is made and was based on the linearization of the model followed by a Pad\'{e} approximation of the exponential terms corresponding to delays. The proof developed above is much more insightful since no approximation is made on the initial model, only assumptions on the network topology. This shows that the static-link model has an application domain which is much wider than the ratio-link and the joint-link models since it is valid in the nonlinear setting and does not result from an approximation and hence is globally valid provided that the assumptions on the network topology are met.

By substituting the above static-link model (\ref{eq:sl}) into the user model (\ref{eq:FAST}) with homogeneous delays, we obtain a new model involving the dynamics of the window sizes only. The local stability of this model has been studied in detail in \cite{Briat:11k}.

\section{Model validation}\label{sec:validation}
The considered scenarios for validating the proposed model are taken from \cite{Jacobsson:08,Tang:10}. The results from NS-2 have been slightly shifted in time so that the congestion window variation times match. Unlike \cite{Tang:10}, the results are not shifted vertically and hence, this results in small discrepancies. If, however, the NS-2 results were shifted vertically so that they match initial equilibrium values, then the curves would match almost perfectly.

\subsection{Single-Buffer/Multiple-Users}

\begin{figure}[H]
  \centering
        \psfrag{b-}[c][c]     {\small{$b^-$}}
        \psfrag{b+}[c][c]     {\small{$b^+$}}
        \psfrag{u1-}[c][c]     {\small{$u_1^-$}}
        \psfrag{u1+}[c][c]     {\small{$u_1^+$}}
        \psfrag{u2-}[c][c]     {\small{$u_2^-$}}
        \psfrag{u2+}[c][c]     {\small{$u_2^+$}}
  \includegraphics[width=0.2\textwidth]{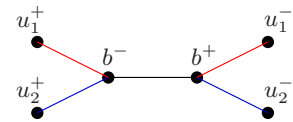}
  \caption{Topology of scenarios 1 \& 2}\label{fig:top1}
\end{figure}

\begin{figure}[H]
\begin{minipage}[b]{0.49\linewidth}
\centering
 \includegraphics[width=\textwidth]{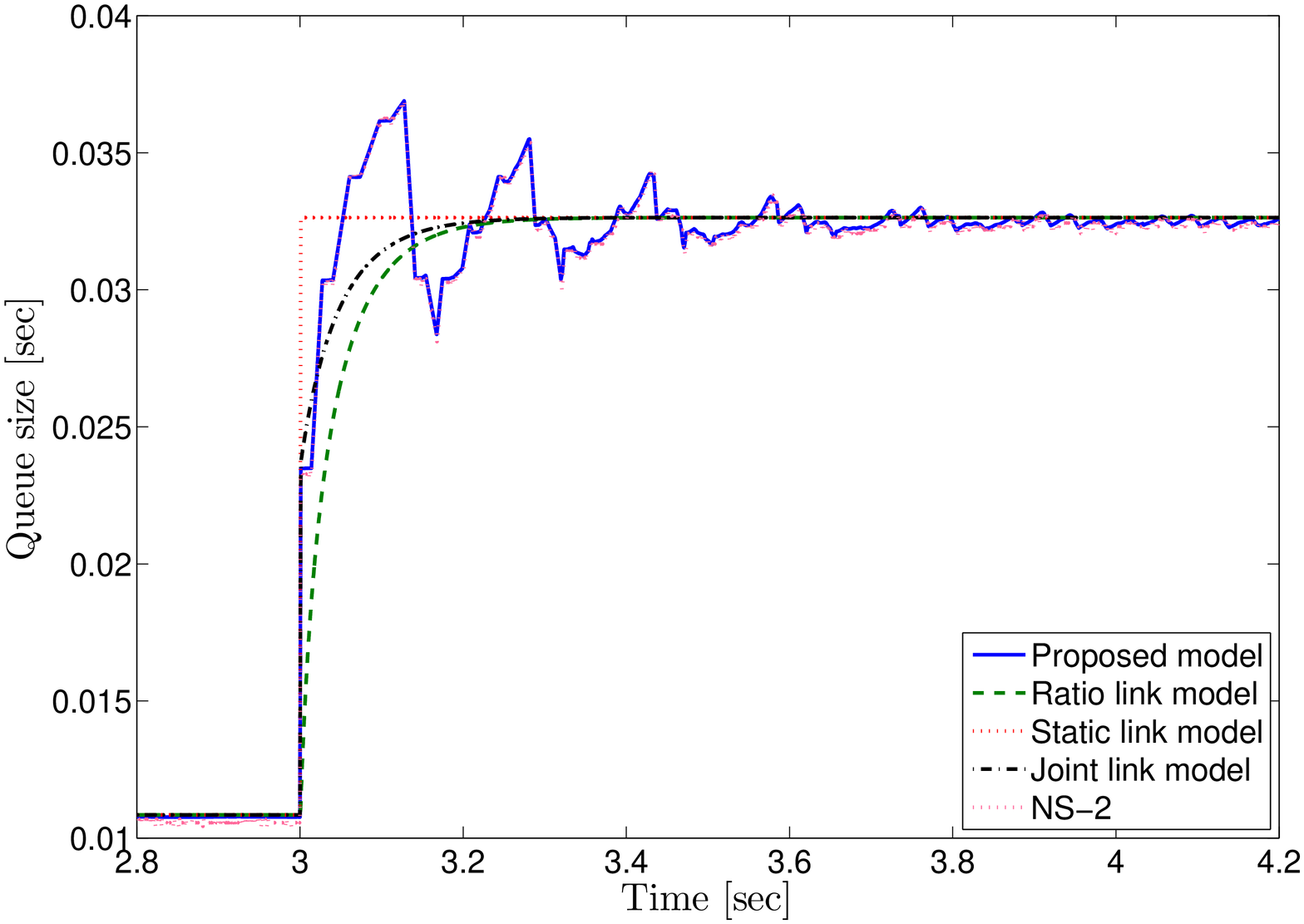}
\caption{Scenario 1: Queue size}\label{fig:ex1}
\end{minipage}
\hfill
\begin{minipage}[b]{0.49\linewidth}
\centering
 \includegraphics[width=\textwidth]{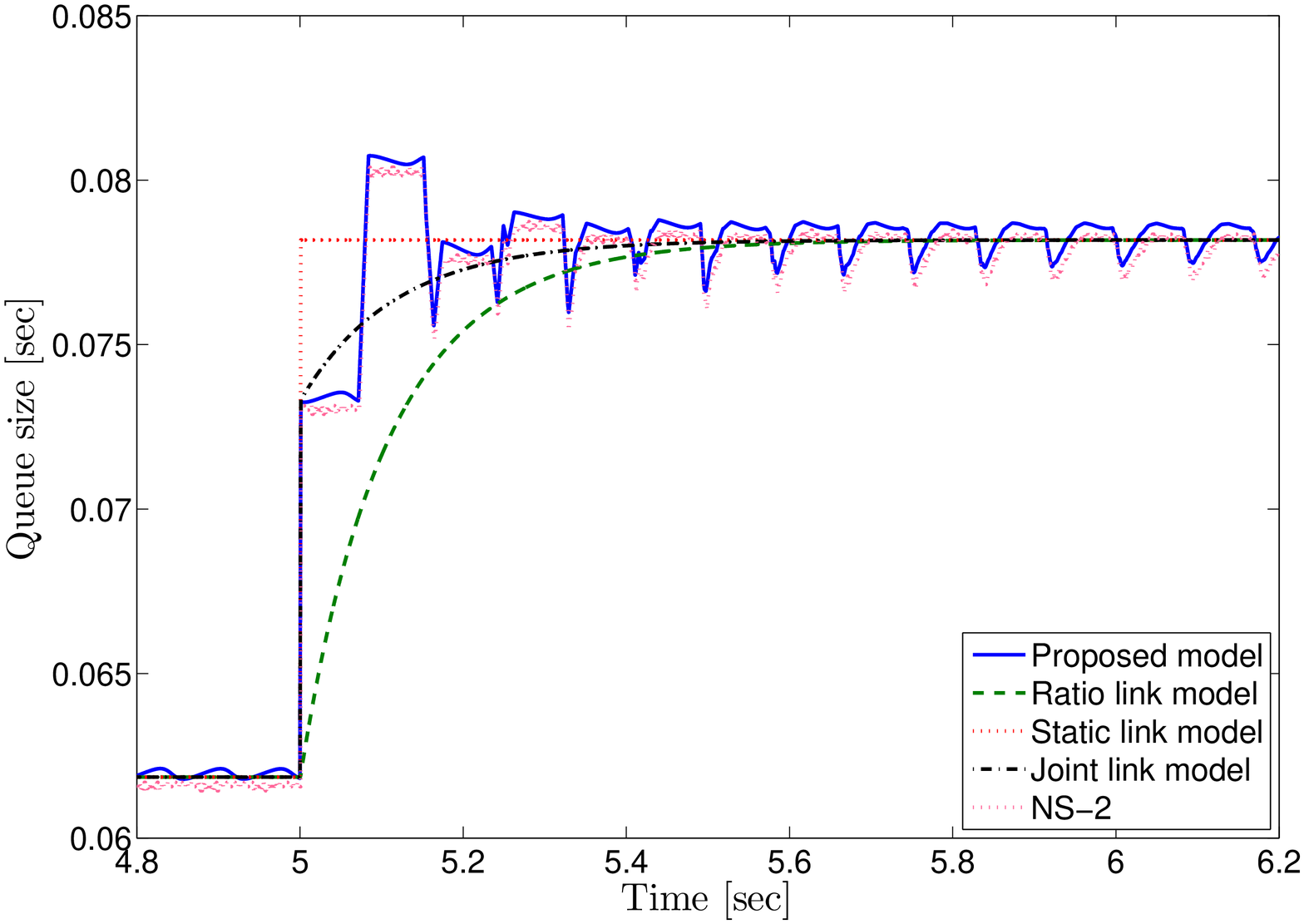}
\caption{Scenario 2: Queue size}\label{fig:ex2}
\end{minipage}
\end{figure}

In this section, we consider the interconnection of two users through a single resource, as depicted in Fig. \ref{fig:top1}. The bottleneck has capacity $c=100$Mb/s and the packet size including headers is 1590 bytes. The following scenarios from \cite{Jacobsson:08,Tang:10} are considered:
\begin{itemize}
  \item Scenario 1: the congestion window sizes are initially $w_1^0=50$ and $w_2^0=550$ packets, at 3s $w_1$ is increased to 150 packets. The propagation delays are $T_1=3.2$ms and $T_2=117$ms for users 1 and 2 respectively; see Fig. \ref{fig:ex1}.
  \item Scenario 2: the congestion window sizes are initially $w^0_1=210$ and $w_2^0=750$ packets, at 5s $w_1$ is increased to 300 packets. The propagation delays are $T_1=10$ms and $T_2=90$ms for users 1 and 2 respectively; see Fig. \ref{fig:ex2}.
\end{itemize}
We can see that the obtained results match quite well NS-2 simulations and are very close to the results reported in \cite{Jacobsson:08,Tang:10}. This is expected since both models are based on conservation laws which are identical when the flight-size is smaller than the congestion window size. Note also that the model captures both the transient phase and the steady-state, unlike the static-, ratio- and joint-flow models.

\subsection{Multiple-Buffers/Multiple-Users}

\begin{figure}[H]
  \centering
        \psfrag{b1-}[c][c]     {\small{$b_1^-$}}
        \psfrag{b1+}[c][c]     {\small{$b_1^+$}}
        \psfrag{b2-}[c][c]     {\small{$b_2^-$}}
        \psfrag{b2+}[c][c]     {\small{$b_2^+$}}
        \psfrag{u1-}[c][c]     {\small{$u_1^-$}}
        \psfrag{u1+}[c][c]     {\small{$u_1^+$}}
        \psfrag{u2-}[c][c]     {\small{$u_2^-$}}
        \psfrag{u2+}[c][c]     {\small{$u_2^+$}}
        \psfrag{u3-}[c][c]     {\small{$u_3^-$}}
        \psfrag{u3+}[c][c]     {\small{$u_3^+$}}
        \psfrag{d-}[c][c]     {\small{$\delta^-$}}
        \psfrag{d+}[c][c]     {\small{$\delta^+$}}
  \includegraphics[width=0.35\textwidth]{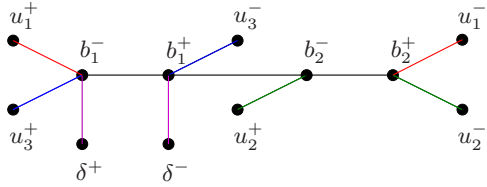}
  \caption{Topology of scenarios 3 to 6 ($\delta$ represents cross-traffic I/O nodes).}\label{fig:top2}
\end{figure}

\begin{figure}[H]
\begin{minipage}[b]{0.49\linewidth}
\centering
 \includegraphics[width=\textwidth]{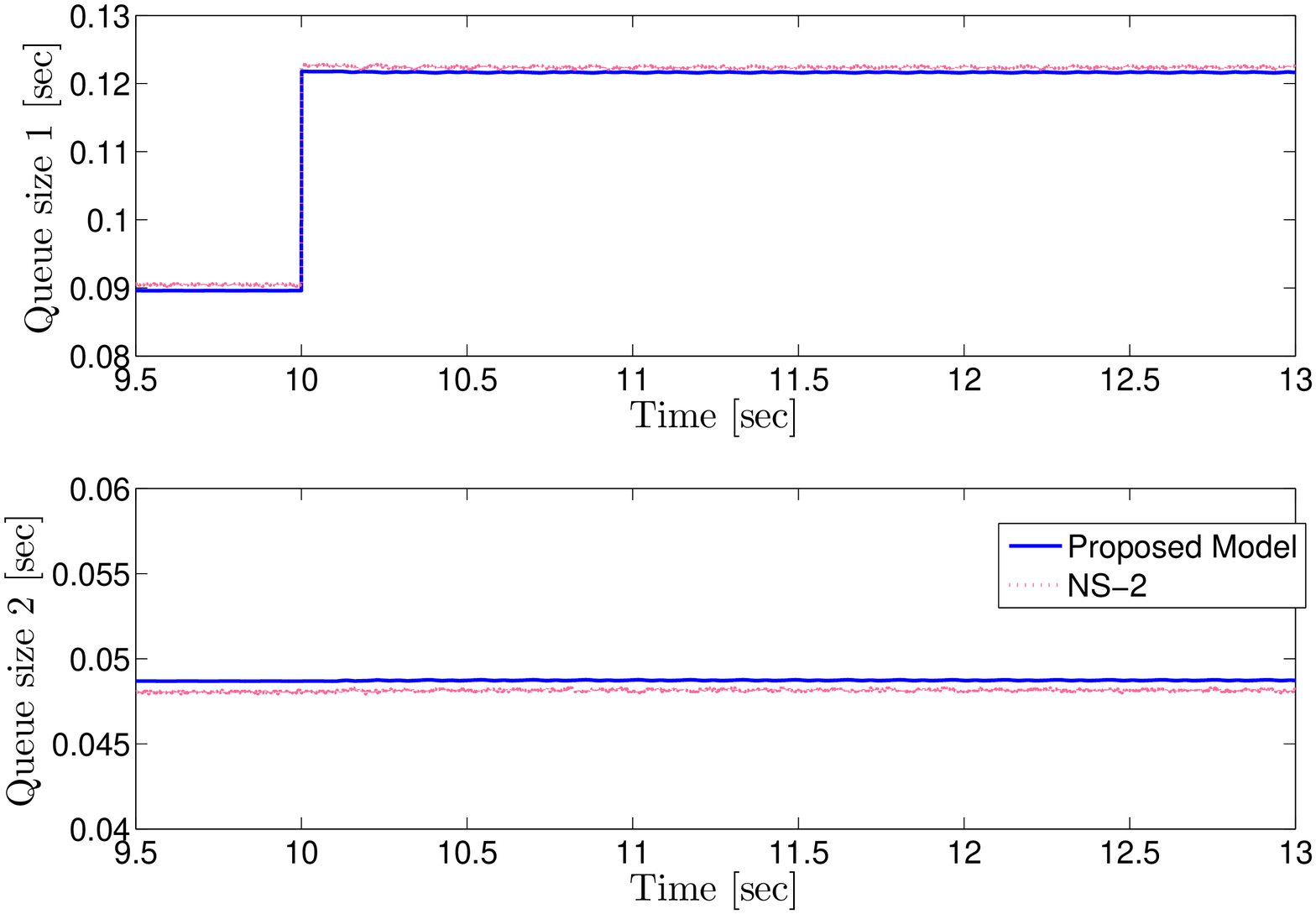}
\caption{Scenario 3: queue 1 (top) and queue 2 (bottom)}\label{fig:ex3}
\end{minipage}
\hfill
\begin{minipage}[b]{0.49\linewidth}
\centering
 \includegraphics[width=\textwidth]{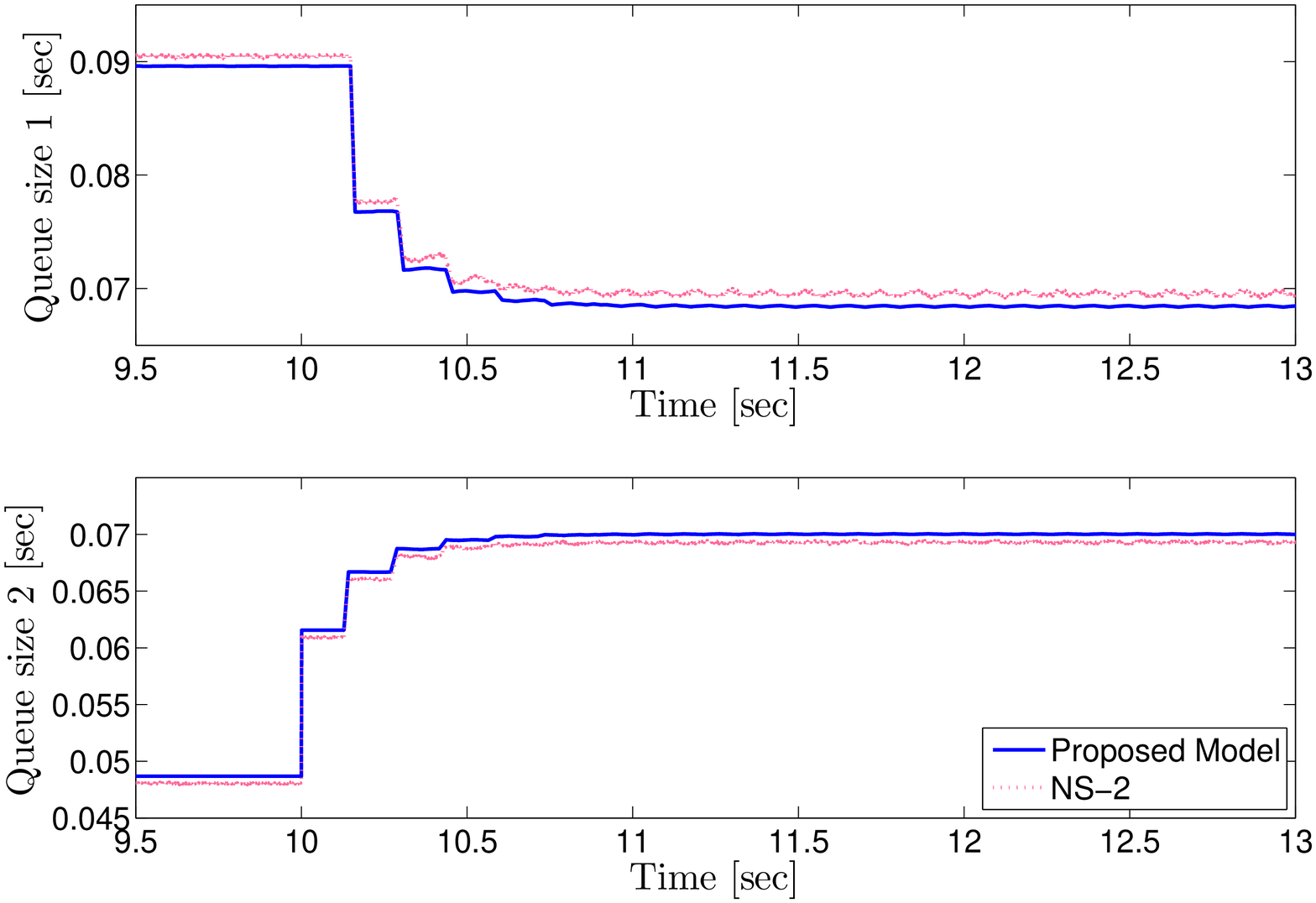}
\caption{Scenario 4: queue 1 (top) and queue 2 (bottom)}\label{fig:ex4}
\end{minipage}
\end{figure}

Here, we consider the case of two buffers interconnected in series (see Fig. \ref{fig:top2}) with capacities $c_1=72$Mb/s and $c_2=180$Mb/s. The packet size including headers is $1448$ bytes. The link propagation delays are $20$ms for link 1 and $40$ms for link 2. The total round-trip propagation delays are $T_1=120$ms, $T_2=80$ms and $T_3=40$ms for sources 1, 2 and 3 respectively. Initially, the congestion window sizes are $w_1^0=1600$ packets, $w_2^0=1200$ packets and $w_3^0=5$ packets. The following scenarios from \cite{Jacobsson:08,Tang:10} are considered:
\begin{itemize}
  \item Scenario 3: No cross-traffic and the congestion window $w_1$ is increased by 200 packets at 10s; see Fig. \ref{fig:ex3}.
  \item Scenario 4: No cross-traffic and the congestion window $w_2$ is increased by 200 packets at 10s; see Fig. \ref{fig:ex4}.
\end{itemize}
Again, the model is able to capture the network behavior well and retrieve previous results given in \cite{Jacobsson:08,Tang:10}. This is due to the equivalence between the ACK-clocking models in this case.

\begin{figure}[H]
\begin{minipage}[b]{0.49\linewidth}
\centering
 \includegraphics[width=\textwidth]{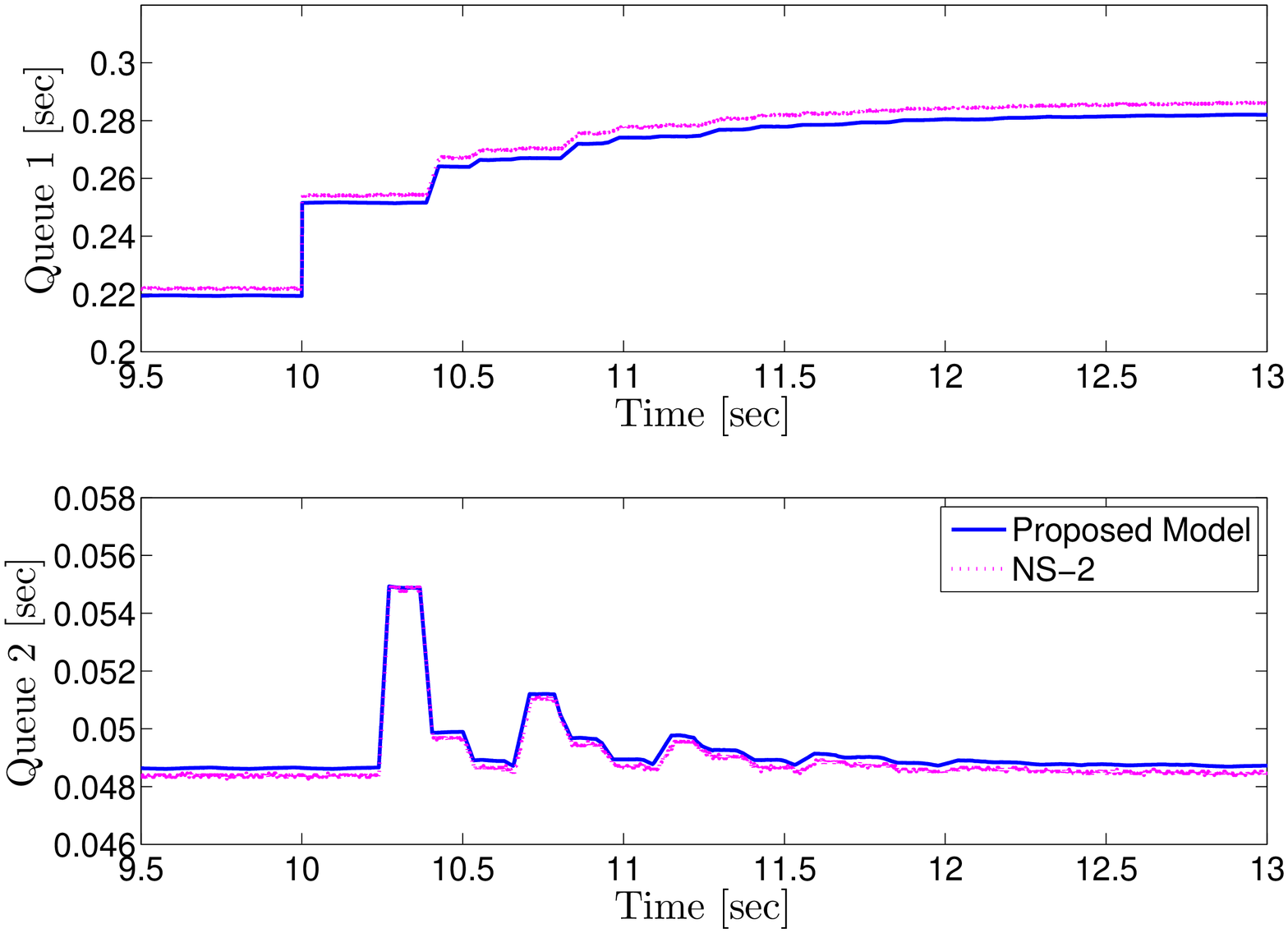}
\caption{Scenario 5: queue 1 (top) and queue 2 (bottom)}\label{fig:ex5}
\end{minipage}
\hfill
\begin{minipage}[b]{0.49\linewidth}
\centering
 \includegraphics[width=\textwidth]{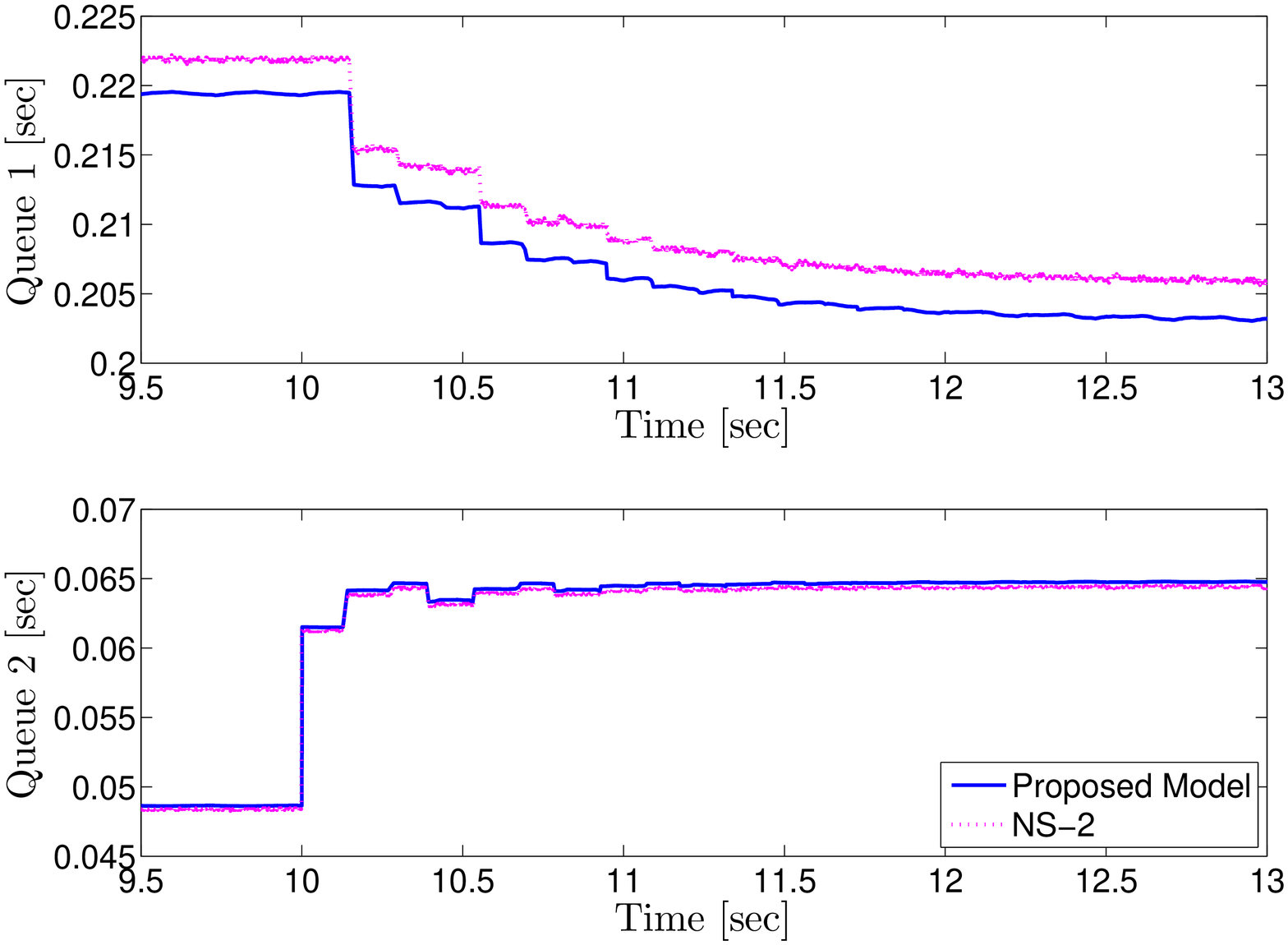}
\caption{Scenario 6: queue 1 (top) and queue 2 (bottom)}\label{fig:ex6}
\end{minipage}
\end{figure}

We introduce now a constant cross-traffic $x_{c1}=c_1/2$ on the first link. Initially\footnote{This scenario is actually identical to the one in \cite[Section III.B.3]{Tang:10}, the congestion window sizes initial values in the paper are inexact.}, we set $w_1^0=1200$, $w_2^0=1600$, $w_3^0=5$ and we consider the following scenarios:
\begin{itemize}
  \item Scenario 5: The congestion window $w_1$ is increased by 200 packets at 10s; see Fig. \ref{fig:ex5}.
  \item Scenario 6: The congestion window $w_2$ is increased by 200 packets at 10s; see Fig. \ref{fig:ex6}.
\end{itemize}

Again the obtained results are identical to the results given in \cite{Jacobsson:08,Tang:10} and match NS-2 simulations. Notice the reaction time, consisting of the sum of the queueing delay and the propagation delay of link 1, between the moment when the congestion window size increases and the moment when the second queue sees the flow increases. This illustrates that the model captures well the communication path and the order of elements.

\subsection{Decrease in Congestion Window Size}
The models proposed in \cite{Zhang:10,Tang:10} do not capture sudden decreases in the congestion window size that would cause the buffer to empty or become smaller than the actual flight-size, that is, smaller than the number of packets in flight. The proposed model captures these phenomena since 1) the ACK-clocking model derived from the conservation law involves the flight-size rather than the congestion window size \cite{Tang:10}; and 2) the user model involves an ACK-buffer to count the number of packets to remove from the network before starting to send again. Note that the derivation of the user model including the ACK-buffer has been possible thanks to the availability of an explicit expression for the flow of acknowledgments, itself made computable through an explicit solution for the queuing delay and the buffer output flows \cite{Briat:10,Briat:11k}. In \cite{Jacobsson:08}, this case is handled by adding a rate limiter to constrain the (negative) slope of the queue size. It is however rather difficult to characterize due to the time-varying nature of the lower-bound on the slope which depends on the received rate of acknowledgment, unfortunately unavailable in the framework of the thesis \cite{Jacobsson:08}.

%Let consider again the two-queue problem with $x_{c1}=c_1/2,x_{c2}=0,w_1^0=1600,w_2^0=1200$. At $t=2$ seconds, the window $w_1$ is halved.
%
%\begin{figure}[H]
%\begin{minipage}[b]{0.49\linewidth}
%\centering
% \includegraphics[width=\textwidth]{./fig/tang/ex10/queues}
%\caption{Queue 1 (top) and queue 2 (bottom)}\label{fig:ex10a}
%\end{minipage}
%\hfill
%\begin{minipage}[b]{0.49\linewidth}
%\centering
% \includegraphics[width=\textwidth]{./fig/tang/ex10/signals}
%\caption{ACK buffer $\pi$ (top) and ACK flow (bottom)}\label{fig:ex10b}
%\end{minipage}
%\end{figure}
%
%We also compare to the experiments and simulations in \cite[Section 5.3]{Jacobsson:08} where a rate limiter is used in order to account for the flight-size dynamics. Recall that in the present model the flight-size dynamics are taken account in the ACK-buffer located at the user level which integrates (absorbs) the ACK-flow when the flight-size has to be decreased.

\begin{figure}[H]
\begin{minipage}[b]{0.49\linewidth}
\centering
 \includegraphics[width=\textwidth]{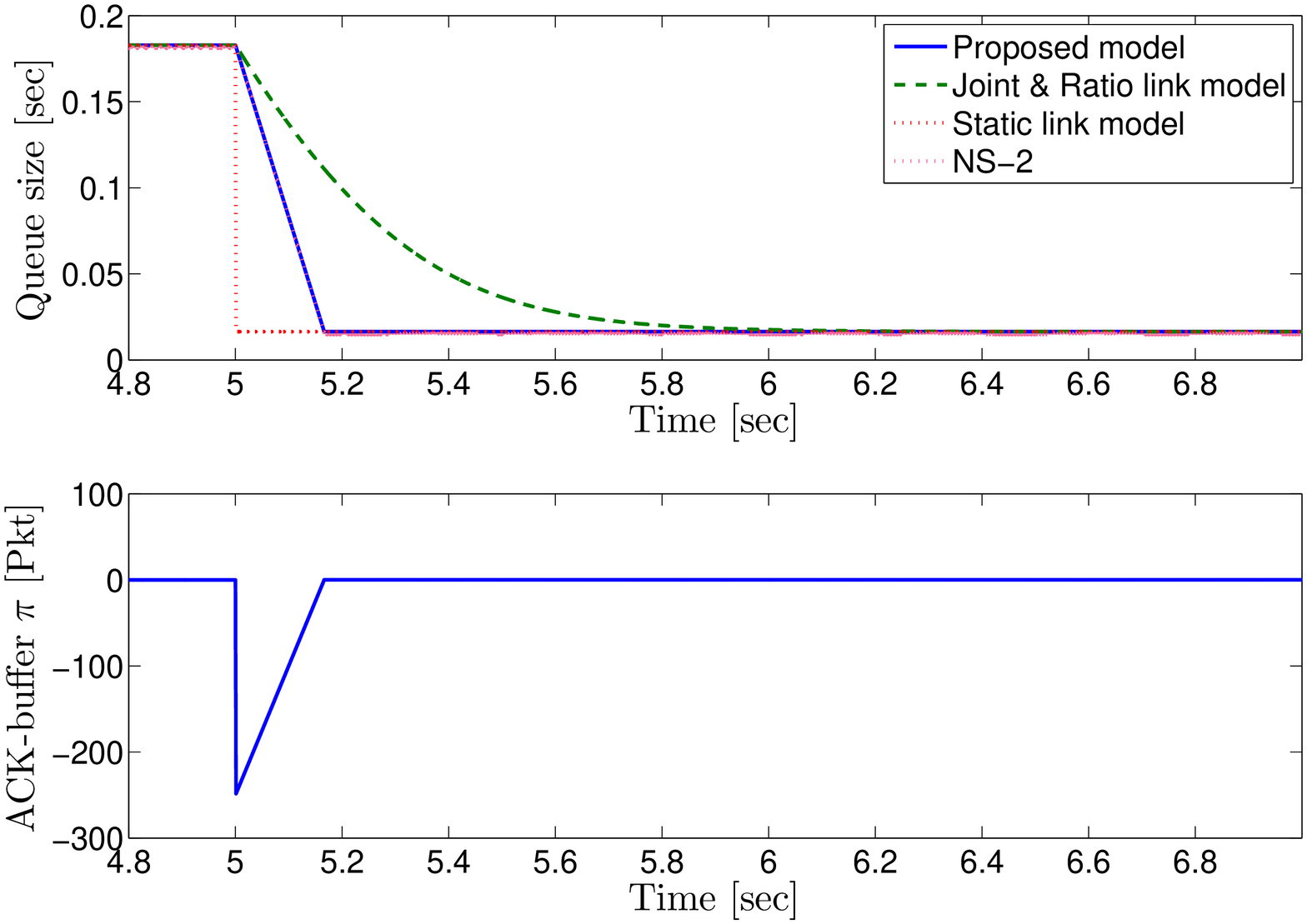}
\caption{Scenario 7: Queue size (top) ACK buffer (bottom)}\label{fig:ex_1k}
\end{minipage}
\hfill
\begin{minipage}[b]{0.49\linewidth}
\centering
  \includegraphics[width=\textwidth]{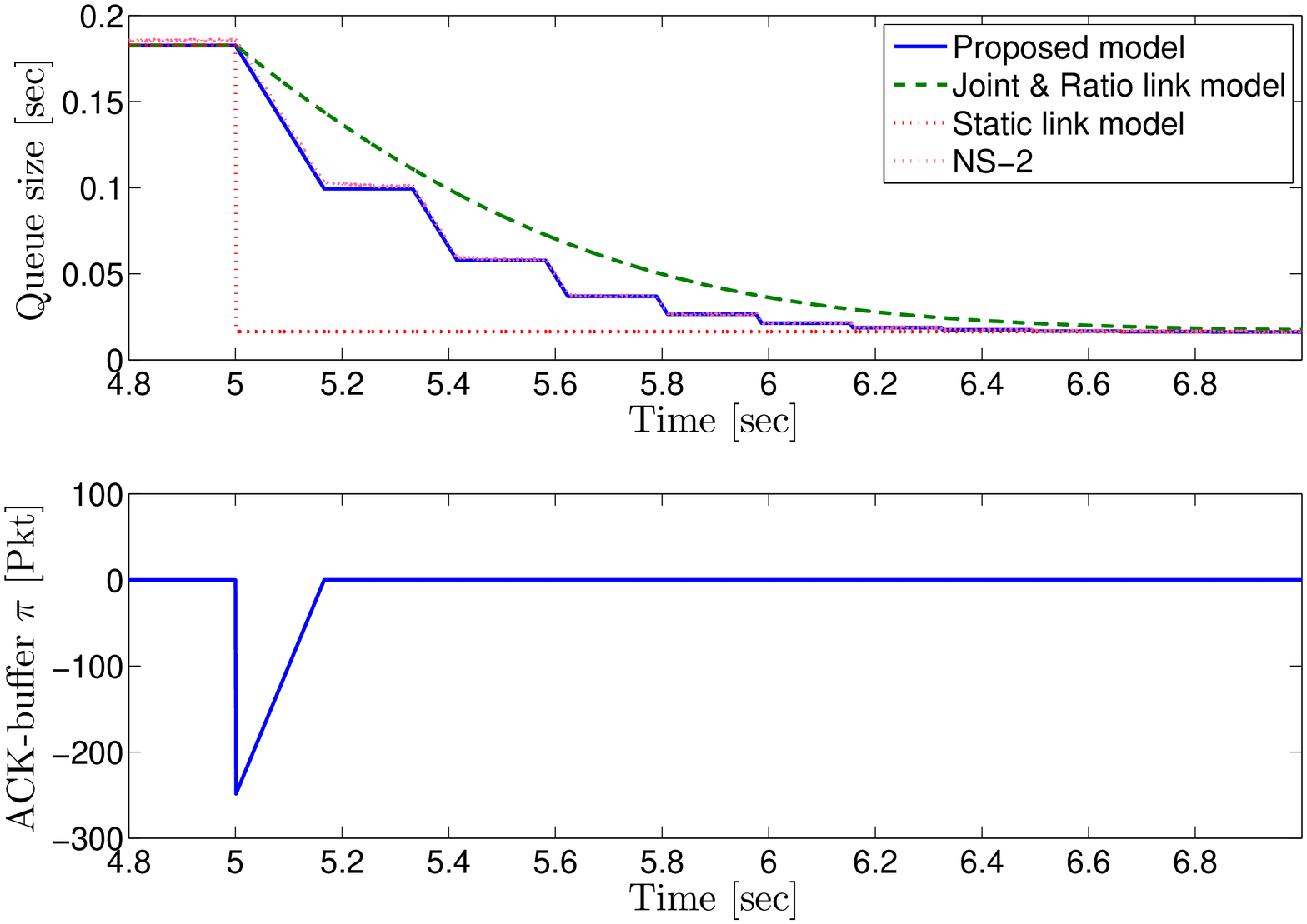}
\caption{Scenario 8: Queue size (top) ACK buffer (bottom)}\label{fig:ex_2k}
\end{minipage}
\end{figure}

Let us consider the single-user/single-buffer case where the total propagation delay is $T=150$ms, the packet size including headers is $1040$ bytes and the initial value of the congestion window size is $w^0=500$. A $t=5$ seconds, the congestion window size is halved. We consider the following scenarios
\begin{itemize}
  \item Scenario 7: $c=12.5$Mb/s and no cross-traffic; see Fig. \ref{fig:ex_1k}.
  \item Scenario 8: $c=25$Mb/s and half capacity used by cross-traffic; see Fig. \ref{fig:ex_2k}.
\end{itemize}
We can see that we obtain exactly the same results as NS-2 simulations (and the rate-limiter model reported in \cite{Jacobsson:08}). As desired, the ACK-buffer measures (counts) the number of packets to remove before starting to send again. 

\section{Conclusion}\label{sec:conclusion}
This paper presents a modular fluid-flow model to analyze congestion in communication networks with arbitrary topology where all data sources use window flow control. Network elements such as queues and sources, are modeled as building blocks using the information conservation law which states that the information is either in transit, lost, or received. The proposed model implements the mechanisms ignored by the previously proposed models, notably at the queue and source levels. It is generic and independent of transport protocol specific congestion control algorithms. Previous models from the literature can be recovered from exact reduction or approximation of this new model. The results obtained from the model match perfectly with the ones obtained from packet-level simulations.

Future works will be devoted to the modeling of data loss, such as packet drops, and the time-out mechanism at the user level.

\section{Acknowledgments}\label{sec:conclusion}
The authors gratefully thank K.-H. Johansson, Ulf. T. J\"{o}nsson, H. Hjalmarsson and H. Sandberg for fruitful discussions.

\bibliographystyle{IEEEtran}

\end{document}

\maketitle

  \setcounter{figure}{5}

  \begin{figure}[H]
\centering
 \includegraphics[width=\textwidth]{./fig/squareflow/flowcompsquare_infocom}
  \caption{Predicted output flows. Plain: output flows, dashed: input flows}\label{fig:compflow}
\end{figure}

\begin{figure}
\centering
 \includegraphics[width=\textwidth]{./fig/squareflow/NS2_comp_SW}
\caption{Model predicted and NS-2 simulated input and output number of packets: flow 1 (top) and flow 2 (bottom).}\label{fig:compflow2}
\end{figure}

\setcounter{figure}{9}

\begin{figure}
\centering
 \includegraphics[width=\textwidth]{./fig/tang/ex1/queue2}
\caption{Scenario 1: Queue size}\label{fig:ex1}
\end{figure}

\begin{figure}
\centering
 \includegraphics[width=\textwidth]{./fig/tang/ex2/queue2}
\caption{Scenario 2: Queue size}\label{fig:ex2}
\end{figure}

\setcounter{figure}{12}

\begin{figure}
\centering
 \includegraphics[width=\textwidth]{./fig/tang/ex3/queues2}
\caption{Scenario 3: queue 1 (top) and queue 2 (bottom)}\label{fig:ex3}
\end{figure}

\begin{figure}
\centering
 \includegraphics[width=\textwidth]{./fig/tang/ex4/queues2}
\caption{Scenario 4: queue 1 (top) and queue 2 (bottom)}\label{fig:ex4}
\end{figure}

\begin{figure}
\centering
 \includegraphics[width=\textwidth]{./fig/tang/ex5/queues2}
\caption{Scenario 5: queue 1 (top) and queue 2 (bottom)}\label{fig:ex5}
\end{figure}
\begin{figure}
\centering
 \includegraphics[width=\textwidth]{./fig/tang/ex6/queues2}
\caption{Scenario 6: queue 1 (top) and queue 2 (bottom)}\label{fig:ex6}
\end{figure}

\begin{figure}
\centering
 \includegraphics[width=\textwidth]{./fig/krister/ex1b/signals2}
\caption{Scenario 7: Queue size (top) ACK buffer (bottom)}\label{fig:ex_1k}
\end{figure}
\begin{figure}
\centering
  \includegraphics[width=\textwidth]{./fig/krister/ex2b/signals2}
\caption{Scenario 8: Queue size (top) ACK buffer (bottom)}\label{fig:ex_2k}
\end{figure}